\documentclass[aps,prd,preprint,groupedaddress,showpacs,nofootinbib]{revtex4}

\usepackage{amsmath}
\usepackage{amssymb}
\usepackage{mathtools}
\usepackage{amsthm}

\theoremstyle{plain}

\usepackage{graphicx}
\usepackage{hyperref}
\usepackage{bbold}
\usepackage{cancel}

\usepackage{enumerate}

\DeclareGraphicsExtensions{.pdf,.png,.jpg,.eps}
\usepackage[utf8]{inputenc}
\usepackage{amsmath}
\begin{document}
\unitlength = 1mm

\title{Is conformal symmetry really anomalous?}

%\title{Does conformal anomaly really exist?}

\author{Stefano Lucat}
\email{s.lucat@uu.nl}

\author{Tomislav~Prokopec}
\email{t.prokopec@uu.nl}

\affiliation{Institute for Theoretical Physics, Spinoza Institute and EMME$\Phi$, Utrecht University,\\
Postbus 80.195, 3508 TD Utrecht, The Netherlands}

\date{\today}

\begin{abstract}

The conformal anomaly (also known as the stress-energy trace anomaly) of an interacting quantum theory,
associated with violation of Weyl (conformal) symmetry by quantum effects, can be amended
if one endows the theory with a dilatation current coupled to a vector field that is the gauge
connection of local Weyl symmetry transformations.
The natural candidate for this Weyl connection is the trace of the geometric torsion tensor, especially if one
recalls that pure (Cartan-Einstein) gravity with torsion is conformal.
We first point out that both canonical and path integral quantisation respect Weyl symmetry.
The only way quantum effects can violate conformal symmetry is by the process of
regularization. However, if one calculates an effective action from a conformally invariant classical theory 
by using a regularisation procedure that is conform with Weyl symmetry, then 
the conformal Ward identities will be satisfied. In this sense Weyl symmetry is not broken by quantum effects.
This work suggests that Weyl symmetry can be treated on equal footing with gauge symmetries and gravity,
for which an infinite set of Ward identities guarantees that they remain unbroken by quantum effects.
\end{abstract}


\maketitle

%%%%%%%%%%%%%%%%%%%%%%%%%%%%%%%%%%%%%%%%%%%%%%%%%%%%%%%%%%%%%%%%%%%%%%%%%
%%%%%%%%%%%%%%%%   I N T R O D U C T I O N   %%%%%%%%%%%%%%%%%%%%%%%%%%%%
%%%%%%%%%%%%%%%%%%%%%%%%%%%%%%%%%%%%%%%%%%%%%%%%%%%%%%%%%%%%%%%%%%%%%%%%%

\section{Introduction}
\label{Introduction}

The discovery of conformal anomaly dates back to 1974 to the seminal work of Capper and Duff~\cite{Capper1974},
in which the authors showed that
the Ward identities of conformal symmetry are broken by the 1-loop quantum fluctuations.
Next Capper and Duff found that the one-loop photon contributes anomalously to the graviton self-energy
through the time-ordered energy-momentum tensor (TT) correlator on Minkowski space,
$\langle T[T_\mu^\mu(x) T_{\sigma\lambda}(0)]\rangle \neq 0$.
Their results show that, while the counter-terms proportional to $1/(D-4)$ respect Weyl symmetry,
where $D$ denotes the dimension of spacetime, the finite contribution does not, yielding an anomalous contribution
to the $TT$ correlator.
In Ref.~\cite{Birrell:1982ix} a second type of anomaly, related to the Euler characteristic of the (Euclidean) manifold,
appeared in the trace anomaly, given in four dimensions by the Gauss-Bonnet
density, namely $\langle T_\mu^\mu\rangle \propto  R^2-4R_{\mu\nu}R^{\mu\nu}+R_{\mu\nu\lambda\sigma}R^{\mu\nu\lambda\sigma}$,
where $R_{\mu\nu\lambda\sigma}$ is the curvature tensor and
$R_{\mu\nu}=g^{\lambda\sigma}R_{\lambda\mu\sigma\nu}$, $R=g^{\mu\nu}R_{\mu\nu}$.

Inspired by these early results, much work has been done on conformal anomalies in the past 40 years.
For example, Ref.~~\cite{Christensen:1977jc} has argued that the anomaly is responsible for Hawking radiation.
Furthermore, the general form of the anomaly in four dimensions was deduced from general covariance and conformal invariance~\cite{Deser:1976yx}, based on which Riegert showed to
follow from a non-local action~\cite{Riegert:1984kt}. More precisely, Riegert showed that all the anomalous
terms -- including $\Box R$, ${\rm Weyl}^2$
and the Gauss-Bonnet term -- follow from variation of the Riegert (nonlocal) action,
where Weyl stands for the Weyl tensor %(the traceless part of the curvature tensor)
 and $\Box$ is the d'Alembertian.
The exception is the anomalous term $\propto R^2$, for which Riegert had shown that it cannot be obtained by
variation of a non-local action.

The conformal anomaly has found many intriguing physical applications. For example,
it has been argued to be the explanation for the dark energy of the
Universe~\cite{Antoniadis:2006wq}.
Furthermore, conformal anomaly may be responsible for formation of cosmological perturbations~\cite{Antoniadis:1996dj},
may induce non-Gaussianties in the cosmic microwave background radiation~\cite{Antoniadis:2011ib}
and could play an important role in the formation of compact stellar objects such as gravastars~\cite{Mottola:2010gp},
which have been proposed as an alternative to black holes.

The common explanation for these results found in literature is that Weyl symmetry may be a symmetry of classical theory,
but it is generally violated in quantum theory.
This can be seen from the non-invariance of the path integral measure~$\mathcal{D}\phi$ under the Weyl transformations,
$g_{\mu\nu}\rightarrow \Omega^2(x) g_{\mu\nu}\,,\, \phi\rightarrow \Omega^{-\frac{D-2}{2}}$~(see for example~\cite{Mukhanov:2007zz}).
%This argument, however, employs Wick rotation to construct the Euclidean action, assuming this is a well defined procedure
%on curved Lorenzian manifolds.
In this work we show that this argument is not true and that both canonical and path integral quantization
(in its phase space form)
are manifestly Weyl invariant thanks to the non-trivial contributions to the path integral measure coming from the gravitational field.
\footnote{In this work we take gravity to be classical (non-dynamical).
The effects of dynamical gravity will be discussed in a separate publication.}

A violation of Weyl symmetry is usually introduced by the procedure of regularization of a (perturbative) effective action.
That is in fact necessary, since usual regularization scheme necessarily introduce a scale in the theory, thus breaking the symmettry.
However, as we will show in section~\ref{Ward identities for Weyl Symmetry} there might be a way of working around this, 
if a compensating field for conformal transformations is present in the theory.

%%%%%%%%%%%%%%%%%%%%%%%%%%%%%%%%%%%%%%%%%%%%%%%%%%%%%%%%%%%%%%%%%%%%%%%%%%%%%%%%%
%%%%%%%%  G R A V I T Y   W I T H   T O R S I O N    %%%%%%%%%%%%%%%%%%%%%
%%%%%%%%%%%%%%%%%%%%%%%%%%%%%%%%%%%%%%%%%%%%%%%%%%%%%%%%%%%%%%%%%%%%%%%%%%%%%%%%%

\section{Gravity with Torsion and its Symmetries}
\label{Gravity with Torsion and its Symmetries}

Recently we have shown~\cite{Lucat:2016eze}
(for reviews see~\cite{Blagojevic:2013xpa,Shapiro:2001rz}) that
geometric torsion can be used to define an exact conformal structure on the manifold
 induced by the transformation laws of the metric tensor $g_{\mu\nu}$ and connection $\Gamma^\lambda{}_{\mu\nu}$,
\begin{equation}
\begin{split}
\label{Conformal.Transformations}
g_{\mu\nu} &\rightarrow \Omega^2 g_{\mu\nu} \,,\\
\Gamma^\lambda{}_{\mu\nu} &\rightarrow \Gamma^\lambda{}_{\mu\nu} + \delta^\lambda_\mu \partial_\nu \log\Omega
\,,
\end{split}
\end{equation}
where $\Omega=\Omega(x)$ is an arbitrary scalar field on the spacetime manifold ${\cal M}$.
The set of local transformations~(\ref{Conformal.Transformations}) requires the introduction of an additional geometric structure on the spacetime manifold, as the transformed connection is not symmetric anymore, and thus generates torsion. In particular, the transformation~(\ref{Conformal.Transformations})
 generates a longitudinal component of the torsion trace one-form $\propto \partial_\mu \log\Omega$. For this
reason we have proposed in~\cite{Lucat:2016eze} to consider the torsion trace as the gauge connection
of localized conformal (Weyl) transformations,~\footnote{As in a large fraction of literature, 
we often refer to Weyl transformations~(\ref{Conformal.Transformations}) as conformal transformations,
even though strictly speaking Weyl transformations constitute just one element of the much larger conformal group,
which consists of the Poincar\'e group augmented by special conformal and Weyl transformations.}
analogously to how the electromagnetic field is the gauge
connection of the localised~$U(1)$ group.

In Ref.~\cite{Lucat:2016eze} we have studied a realization of such a theory, where
by using the metric formalism of spacetime with torsion
we defined the covariant derivative acting on an arbitrary representation of the Lorentz group with conformal weight $\omega$ as,
\begin{equation}
\label{nabla.bar}
 \bar{\nabla}{}_\mu \Psi = \nabla{}_\mu \Psi + (\omega_g - \omega) \mathcal{T}_\mu \Psi\,,
\end{equation}
where $\omega_g$ is the geometric weight (scaling dimension) of $\Psi$,
namely its scaling under the diffeomorphisms, $x^\mu\rightarrow x^\mu + \lambda x^\mu$, and $\omega$ is the scaling dimension of the field. $\nabla{}_\mu$ is the spacetime covariant derivative with torsion, whose connection one can solve for using the metric compatibility condition, to get, when $\nabla{}_\mu$ acts on a vector field,
\begin{equation}
\begin{split}
\nabla{}_\mu V^\lambda &=  \partial_\mu V^\lambda  + \Gamma^\lambda{}_{\sigma\mu} V^\sigma = \partial_\mu V^\lambda + \big\{^\lambda{}_{\sigma\mu}\big\} V^\sigma + K^\lambda{}_{\sigma\mu} V^\sigma\,,\\
\big\{^\lambda{}_{\sigma\mu}\big\} &= \frac{g^{\lambda\rho}}{2}\left (\partial_\sigma g_{\mu\rho} + \partial_\mu g_{\sigma\rho} - \partial_\rho g_{\mu\nu}\right ) \,,\\
K_{\lambda\sigma\mu} &= T_{\lambda\sigma\mu} + T_{\sigma\mu\lambda} + T_{\mu\sigma\lambda}\,\,\&\,\, T^\lambda{}_{\sigma\mu} \equiv \Gamma^\lambda{}_{[\sigma\mu]}\,.
\end{split}
\end{equation}
The derivative operator $\bar{\nabla}$ in~(\ref{nabla.bar}) denotes  
conformal covariant derivative which commutes with
the transformations~(\ref{Conformal.Transformations}), $\bar{\nabla}_\mu \Omega = \Omega \bar{\nabla}_\mu$.
Sometimes we also use the notation $\overset{\circ}{\nabla}_\mu$ to indicate the general relativistic covariant derivative, that is the derivative computed by using Christoffel symbols. 

The conformal covariant derivative in Eq.~(\ref{nabla.bar}) satisfies metric compatibility and conservation of the volume form, since in $D=4$ spacetime dimensions,
$\bar{\nabla}_{\mu} \epsilon^{\alpha\beta\gamma\delta} = 0$.~\footnote{A generalization to general $D$
spacetime dimensions is straightforward, $\bar{\nabla}_{\mu} \epsilon^{\alpha_1\alpha_2\dots \alpha_D} = 0$.}
 Finally~(\ref{nabla.bar}) satisfies Stokes theorem if the integral is dimensionless
({\it i.e.} if its scaling dimension is {\it zero}), namely,
\begin{equation}
\label{stokes.theorem.2} \int_\Sigma d\omega =  \int_\Sigma\,\bar{\nabla}_{\mu_1} \omega_{\mu_2\cdots\mu_p} \text{d}x^{\mu_1}\wedge\cdots\wedge\text{d}x^{\mu_p} =\int_{\partial\Sigma}  \omega_{\mu_2\cdots\mu_p} \text{d}x^{\mu_2}\wedge\cdots\wedge\text{d}x^{\mu_p}\,  \,,
\end{equation}
if the scaling dimension of $\omega_{\mu_2\cdots\mu_p}$ is equal to $0$. This is easily verified using that the factor appearing in the definition of the conformal covariant derivative~(\ref{nabla.bar}) $(\omega_g-\omega)$ would be $+p$.

%%%%%%%%%%%%%%%%%%%%%%%%%%%%%%%%%%%%%%%%%%%%%%%%%%%%%%%%%%%%%%%%%%%%%%%%%%%%%%%%%%%%%%%%%%%%%%%
%%%%%%%%%%   W E Y L   S Y M M E T R Y   I N   T H E  Q U A N T U M   T H E O R Y %%%%%%%%%%%%%
%%%%%%%%%%%%%%%%%%%%%%%%%%%%%%%%%%%%%%%%%%%%%%%%%%%%%%%%%%%%%%%%%%%%%%%%%%%%%%%%%%%%%%%%%%%%%%%

\section{Weyl symmetry in the quantum theory}
\label{Weyl symmetry in the quantum theory}

We begin this section by showing why neither canonical quantisation nor path integral quantisation
can break local Weyl symmetry and how this becomes apparent when coupling to (non-dynamical) gravity is accounted for.

Canonical quantisation in curved spaces requires the existence of a time-like vector, $n^\mu$,
along which the canonical momentum is defined.
This in turn induces a foliation on spacetime~\cite{Birrell:1982ix} with spatial slices, $\Sigma$, on which one can rigorously define field quantisation which is, as we shall now show, Weyl invariant on curved spacetimes.

%\subsection{Scalars and fermions}
%\label{Scalars and fermions}

Let us begin by considering a conformally coupled scalar field $\phi(x)$ whose
classical action is,~\footnote{The action~(\ref{Scalar.Action.1}) is conformal in arbitrary $D$ spacetime dimensions. The quantization
procedure would go through for a self-interacting scalar field in $D=4$, in which the interaction term
$S_{\rm int}=\int d^4x\sqrt{-g}{\cal L}_{\rm int}=-\int d^4x\sqrt{-g}(\lambda/4)\phi^4$. }
\begin{eqnarray}
\label{Scalar.Action.1}
S_\phi &=& \int\text{d}^Dx \sqrt{-g}
\bigg\{\!-\!\frac12 g^{\mu\nu} \left (\partial_\mu \phi
 +\frac{D-2}{2}{\cal T}_\mu\phi\right ) \left (\partial_\nu \phi
+\frac{D-2}{2}{\cal T}_\nu\phi\right )  %\!-\! \frac{\lambda}{4!}\phi^4
\bigg\} \\
&=&\int\text{d}^Dx \sqrt{-g}\bigg\{\!-\!\frac12
 \left [\left (g^{\mu\nu} - \frac{n^\mu n^\nu}{\|n\|^2}\right ) \bar{\nabla}{}^\perp_\mu\phi\bar{\nabla}{}^\perp_\nu\phi + \frac{n^\mu n^\nu}{\|n\|^2}  \bar{\nabla}{}^\parallel_\mu\phi\bar{\nabla}{}^\parallel_\nu\phi\right ]
% \!-\! \frac{\lambda}{4!}\phi^4
\bigg\}
\,,
\end{eqnarray}
where ${\cal T}_\mu$ is the torsion trace 1-form (see Ref.~\cite{Lucat:2016eze}),
our metric convention is ${\rm sign}[g_{\mu\nu}]=(-1,1,1,\dots)$
and
\begin{equation}
 \phi\rightarrow \tilde\phi=\Omega^{-\frac{D-2}{2}}\phi
\,.
\label{canonical dimension phi}
\end{equation}
Then the usual definition of canonical momentum implies,
\begin{eqnarray}
\label{canonical.momentum}
\pi_\phi \equiv \frac{\delta S_\phi}{\delta(n^\nu\partial_\nu\phi) }
  = \sqrt{-g}\,\frac{ n_\nu}{\|n\|^2} g^{\mu\nu}\bar{\nabla}_\mu \phi
\,,\,\implies \pi_\phi\rightarrow\tilde{\pi}_\phi = \Omega^{(D-2)/2} \pi_\phi
\,,
\end{eqnarray}
where ${\|n\|^2}=g(n,n)=g_{\mu\nu}n^\mu n^\nu$
is the norm-squared of $n^\mu$, whose canonical dimension is $2$
(if $t={\rm constant}$ on $\Sigma$ then $n^\mu=\delta^\mu_{\;0}$)
 and
$n^\mu$ does not scale (its scaling dimension is zero).
 Eqs.~(\ref{canonical dimension phi}--\ref{canonical.momentum})
in turn imply,~\footnote{Throughout this work we work in natural units
in which, $\hbar=1=c$ and $\mathbb{1}$ in~(\ref{commutation.relations}) is a shorthand for
$\delta^{D-1}(\vec x-\vec x^\prime)$ times a Kronecker delta over internal indices, if the field contains more than one
component.}
\begin{equation}
\label{commutation.relations}
\big[\phi, \,\pi_\phi\big] = \big[\tilde{\phi},\,\tilde{\pi}_\phi\big] = i{\mathbb{1}}
\,,
\end{equation}
such that canonical quantisation respects conformal symmetry.
This invariance is a simple consequence of the fact the canonically conjugate variables have opposite
scaling dimension, see~(\ref{canonical dimension phi}--\ref{canonical.momentum}).

One can easily show that the same is true for a fermionic field $\psi$. To see that consider
a fermion whose action is,
\begin{equation}
\label{fermion.action}
S_\psi = \int \text{d}^D x \sqrt{-g}
\Big[\, \frac{i}{2} e_a^\mu\left (\bar{\psi} \gamma^a \overset{\rightarrow}{\nabla}{}_\mu \psi - \bar{\psi} \overset{\leftarrow}{\nabla}{}_\mu\gamma^a  \psi \right)+{\cal L}_{\psi,\rm int}\Big]
\,,
\end{equation}
where ${\cal L}_{\psi,\rm int}$ is some (conformal) interaction term (which may include coupling to other matter fields but contains
no derivatives),
$e^\mu_a(x)$ is a tetrad field, $\gamma^a$ are the Dirac matrices on tangent space,
$\{\gamma^a,\gamma^b\}=2\eta^{ab}$, on which the metric is flat Minkowski, $\eta^{ab}={\rm diag}(-1,1,..)$ and $\bar\psi=\psi^\dag\gamma^0$.
The action~(\ref{fermion.action}) then implies a canonical momentum,
\begin{equation}
\label{canonical.momentum.fermions}\pi_\psi = \sqrt{-g}\, \frac{n_\mu}{\|n\|^2} e_a^\mu\bar{\psi}\gamma^a\,,\,\implies \pi_\psi\rightarrow\tilde{\pi}_\psi = \Omega^{(D-1)/2} \pi_\psi\,.
\end{equation}
Taking into account Fermi statistic, the quantisation condition for fermions becomes,
\begin{equation}
\label{commutation.relations.fermions}
\big \{\psi, \,\pi_\psi\big \}=\big \{\tilde{\psi}, \,\tilde{\pi}_\psi\big \} = \mathbb{1}
\,,
\end{equation}
proving that canonical quantisation of fermions respects conformal symmetry.

Next we consider conformal symmetry in the context of path integral quantisation. 
It turns out that the symmetry of quantisation is made manifest in the phase space version of path integral,
which is usually taken to define the path integral quantisation.

For an interacting scalar field theory, for example, the vacuum-to-vacuum scattering amplitude
reads,~\footnote{For fermions the measure is,
\[\mathcal{D}\psi\mathcal{D}\pi_\psi =\mathcal{D}\psi\mathcal{D}\bar{\psi}\det\left (\sqrt{-g} \|n\|^{-2} n^\nu g_{\nu\mu} \gamma^\mu\right ),\]
where the determinant is taken both on spinor indices and on spacetime continuous indices. The measure is both diffeomorphism
 and Weyl invariant.}
\begin{equation}
\label{pathintegral}
\langle {\it in}|{\it out}\rangle
= \int \mathcal{D}\phi\mathcal{D}\pi_\phi \exp\left\{i\int \text{d}^{D-1}\vec{x}\text{d}t
\big(\pi_\phi n^\mu \partial_\mu\phi - {\cal H}_\phi\big)\right\}\,,
\end{equation}
where $n^\mu \partial_\mu\phi = \dot{\phi}$ if the spatial hypersurface $\Sigma$ is
chosen to be a constant time hypersurface, 
${\cal H}_\phi(\phi,\pi_\phi)=\pi_\phi n^\mu\partial_\mu\phi- \sqrt{-g}{\cal L}_\phi(\phi,\partial_\mu\phi)$ 
is the Hamiltonian density and ${\cal L}_\phi$ is the Lagrangian density
(which for now needs not be specified).
With Eqs.~(\ref{canonical dimension phi}) and~(\ref{canonical.momentum})
in mind we immediately see that the measure in~(\ref{pathintegral}) is Weyl invariant, such that the path integral
in~(\ref{pathintegral}) and thus also the scattering amplitude
must be Weyl invariant if ${\cal L}_\phi$ is conformal.

The only thorny issue that might spoil conformal symmetry is related to the question of whether the path integral~(\ref{pathintegral}) is well defined.
That indeed may pose a problem in the sense that the amplitude~(\ref{pathintegral}) is generally divergent and
since any regularisation of~(\ref{pathintegral}) violates Weyl symmetry, it can make it `anomalous.'
However, as we argue below, a suitable regularisation scheme can make Weyl symmetry non-anomalous.

It is worth remarking that in literature one often finds a path integral formulation in which the integration over the momentum
is performed and in which Weyl symmetry of the path integral does not seem manifest.
To show that this is not the case, let us  perform
the Gaussian integral over the (suitably shifted) momentum,
\begin{equation}
\label{momentum.integrals} 
\int\mathcal{D}\tilde\pi_\phi \exp\left\{i\int\text{d}^Dx \frac{\| n\|^{2}\tilde\pi_\phi^2}{\sqrt{-g}} \right\}
 = \sqrt{\det\left(\sqrt{-g}\| n\|^{-2}\delta^D(x-y)\right )} = \prod\limits_x\left( \frac{\sqrt{-g(x)}}{\| n(x)\|^2}\right )^{\frac{1}{2}}
\,.
\end{equation}
With this result in mind, Eq.~(\ref{pathintegral}) can be written as,
\begin{equation}
\label{pathintegral.momentum.performed}
\langle {\it in}|{\it out}\rangle= \int \bar{\mathcal{D}}\phi \,e^{iS_\phi}\,,
\end{equation}
where $S_\phi=\int d^Dx\sqrt{-g}{\cal L}_\phi$ and 
the barred measure is
\begin{equation}
\bar{\mathcal{D}}\phi = \prod\limits_x d\phi(x)\left( \frac{\sqrt{-g(x)}}{\| n(x)\|^2}\right )^{\frac{1}{2}}
\,,
\label{invariant measure}
\end{equation}
which is obviously Weyl invariant. Note the dependence on the metric tensor in~(\ref{invariant measure}),
which is usually omitted from the measure, but is essential for Weyl symmetry.

For systems with constraints -- such as gauge theories or gravity -- one can also show that
the phase space path integral measure is conformal. Since a
proper analysis of that question is rather subtle, we relegate the details on how
that works in an Abelian gauge theory to Appendix~A. The discussion of gravity 
and non-Abelian gauge theories we postpone to future work.

In this section we have presented cogent arguments in favour of preservation of Weyl symmetry.
Namely, since both canonical (or Dirac) and path integral quantisation preserve conformal symmetry,
if Weyl (conformal) symmetry is respected by classical theory,
it will remain symmetry of the quantum theory provided one uses regularisation scheme that 
does not violate the symmetry. This is to be contrasted with the literature
on conformal anomalies, which states that conformal symmetry is anomalous in
the sense that quantum effects generically break it.

In the next section we proceed with further building evidence and show that,
if a classical theory is conformal, the conformal Ward identities -- when suitably modified --
are obeyed in the quantum theory.

%%%%%%%%%%%%%%%%%%%%%%%%%%%%%%%%%%%%%%%%%%%%%%%%%%%%%%%%%%%%%%%%%%%%%%%%%%%%%%%%%%%%%%%%%%%%%%%
%%%%%%%%%   W A R D   I D E N T I T I E S   F O R   W E Y L   S Y M M E T R Y   %%%%%%%%%%%%%
%%%%%%%%%%%%%%%%%%%%%%%%%%%%%%%%%%%%%%%%%%%%%%%%%%%%%%%%%%%%%%%%%%%%%%%%%%%%%%%%%%%%%%%%%%%%%%%

\section{Ward identities for Weyl Symmetry}
\label{Ward identities for Weyl Symmetry}

This is the principal section of this paper in which we address the central question: ``What is the origin of conformal anomalies?"
\footnote{Conformal anomalies is the commonly used term signifying any anomaly associated with the breaking of Weyl symmetry.}
If we accept the argument of the previous section -- according to which quantisation does not break conformal symmetry -- 
then conformal symmetry should be a symmetry of the quantum theory, at least if the classical action is itself Weyl invariant.
Instead, the literature claims that conformal symmetry is broken by quantisation, as it is 
corroborated/evidenced by breaking of the conformal Ward identity, $\langle T_\mu^\mu\rangle = 0$.
In what follows we show that the problem of conformal anomalies can be solved in theories
in which a compensating field for Weyl symmetry is added. 
The claimed breaking of the quantum identity, $\langle T_\mu^\mu\rangle=0$,
should be re-interpreted as breaking of the global (rescaling) symmetry, while the local symmetry is left unbroken.

\subsection{Fundamental Ward identity}
\label{Fundamental Ward identity}

The idea we want to pursue here is to consider Weyl symmetry as a gauge transformation,
which is compensated by a one-form (Weyl) field which transforms as, $\mathcal{T}_\mu \rightarrow \mathcal{T}_\mu + \partial_\mu \log\Omega(x)$, under Weyl transformations defined 
in~(\ref{Conformal.Transformations}).
%where the Weyl transformation is defined by $g_{\mu\nu}\rightarrow \Omega^2 g_{\mu\nu}$.
As we show below, this then instigates the following modification of the fundamental Ward identity for Weyl symmetry,
\begin{equation}
\label{conformal.symmetry.Ward.identity}
\langle T_\mu^\mu \rangle + \langle \bar{\nabla}{}_\mu \Pi^\mu\rangle=0\,,
\end{equation}
where $\Pi^\mu$ is the dilatation current that sources the Weyl field $\mathcal{T}_\mu$ and the brackets denote the time ordered product of operators. Equation~(\ref{conformal.symmetry.Ward.identity}) is the fundamental Ward identity
for local Weyl symmetry. The identity simply states that there exists a current
$\Pi^\mu$ whose divergence equals to the trace of the energy-momentum tensor.

If a theory is globally scale invariant, we would be led to the stronger requirement that $\langle T_\mu^\mu\rangle = 0$~(since
for global scale transformations, $ \partial_\mu \log\Omega= 0$).
In such a case, at least for flat spacetimes, there exists a conserved current, the dilatation current, which is conserved, namely,
$D^\mu = - T^\mu_{\;\nu} x^\nu$.
From these observations it then follows that requiring $\langle T_\mu^\mu\rangle = 0$ is equivalent to demanding that
the global scale transformation is a symmetry of the theory, which is not the case if the symmetry is {\it e.g.}
 broken by quantum effects.
In other words, one can try to construct a classical action by
using only the metric tensor and matter fields that is Weyl invariant. In constructing such a theory, however, 
one usually makes no distinction between global and local coformal symmetry
and a breaking of global scale symmetry implies a breaking of local conformal symmetry.

The crucial observation is that in flat space there {\it always} exists a dilatation 
current $D^\mu$ such that, 
\begin{equation}
\label{Flat.Space.Limit.Conf.Ward.Identity}\partial_\mu D^\mu = - T_\mu^\mu\,,
\end{equation}
which is divergence-free only if global scale symmetry is realised.
Our proposal is to elevate the current $D^\mu$ to the source for the Weyl gauge field $\mathcal{T}_\mu$ on general curved spacetimes.
If such a Weyl field exists it could be used to generate
the source current {\it via}, $\Pi^\mu = (-g)^{-1/2}\delta S/\delta \mathcal{T}_\mu$.
Hence the physical meaning of $\Pi^\mu$ is the curved spacetime generalisation of the dilatation current $D^\mu$.
Such a current is in general independent of the energy-momentum tensor and moreover -- as we shall see --
can be written as a local function of the fields.~\footnote{The nonlocal expressions,
 $\Pi^\mu(x) = -\int_{x^{(0)}}^{x} d\tilde x^\nu T^\mu_{\;\nu}(\tilde x)$ and
$\Pi^\mu(x) = -(\partial^\mu/\Box)T^\alpha_{\;\alpha}(x)$, would  obviously do. However,
such forms for $\Pi^\mu$ would be obtained by variation of the corresponding nonlocal effective actions.
One could make these actions local by introducing an auxiliary field, whose physical meaning is that of a Weyl field
$\mathcal{T}_\mu$. We may as well bypass the nonlocal step and
from the very beginning work with a local formulation in which $\mathcal{T}_\mu$ exists as an independent field. That is the
approach advocated here.
}
 That fact of Nature seems to hint at the existence of a new symmetry and
it would be foolish not to make use of it.

As we will see next, there are several operators for which $\Pi^\mu$ is non trivial, for example all dimension four curvature operators with torsion
and the scalar field kinetic terms, as in~(\ref{Scalar.Action.1}). All these contributions can get sourced by
a non vanishing energy-momentum tensor trace,
such to respect the identity~(\ref{conformal.symmetry.Ward.identity}).

In order to see that the dilatation current
naturally arises and that it can be written as a local function of the fields, let us consider
an interacting, scale-invariant field theory (in $D=4$) of $N$ scalar fields,
\begin{equation}
\label{O(N).invariant.theory}
S_{\{\phi^a\},N}
=
\int\text{d}^4 x\sqrt{-g} \left ( -\frac{1}{2} \zeta_{ab} \partial_\mu \phi^a \partial^\mu \phi^b
 + \frac{\lambda_{abcd}}{4} \phi^a\phi^b\phi^c\phi^d + \frac{\xi_{ab}}{2} \phi^a\phi^b R\right)
\,,
\end{equation}
where $\zeta_{ab}$, $\xi_{ab}$ and $\lambda_{abcd}$ are constants. 
It is easy to show that the trace of the energy-momentum tensor,
 $T_{\mu\nu}=\frac{2}{\sqrt{-g}}\frac{\delta S}{\delta g^{\mu\nu}}$, satisfies,
\begin{equation}
\label{O(N).energy.momentum.trace}
T_\mu^\mu = \nabla_\mu \left [\left (\zeta_{ab} + 12 \xi_{ab}\right ) \phi^a\partial^\mu \phi^b\right ]
\implies
D^\mu = -\left (\zeta_{ab} + 12 \xi_{ab}\right ) \phi^a\partial^\mu \phi^b
\,.
\end{equation}
Hence our prescription for the dilatation source, namely
$D^\mu = (-g)^{-1/2}\delta S/\delta \mathcal{T}_\mu$, yields naturally to the modification,
$\partial_\mu \rightarrow \bar{\nabla}{}_{\mu}\,,\, R\rightarrow \bar{R}$,
in the action~(\ref{O(N).invariant.theory}),
where $\bar{\nabla}{}_{\mu}$
is the conformal covariant derivative and $\bar{R}$ is the curvature scalar with torsion.
We are then led to the action~(\ref{Scalar.Action.1})
generalised to $N$ interacting scalars with non-minimal coupling to the curvature scalar and quartic interactions.
Then for all values of $\zeta_{ab}$ and $\xi_{ab}$ ($a,b,=1,\dots,N$)
we would have a Weyl invariant action whose energy momentum tensor is the divergence of a vector current.

In Ref.~\cite{Lucat:2016eze} we showed that the natural candidate for the gauge field of Weyl transformations is torsion trace. Indeed,  torsion trace generates scale transformations on vectors that are parallel transported on the manifold
and, if the expected transformations
$g_{\mu\nu}\rightarrow \Omega^2 g_{\mu\nu}\,,\,\mathcal{T}_\mu \rightarrow \mathcal{T}_\mu + \partial_\mu \log \Omega$ are performed,
one finds that the curvature tensor and the geodesic equation are left invariant.
Moreover, since torsion trace is a geometric field, it couples universally to all matter fields.
For all these reasons we conclude that the torsion trace can be considered as the Weyl gauge field,
and that is what we propose in this paper.

In what follows we consider an interacting, conformal scalar theory and prove that the fundamental Ward identity~(\ref{conformal.symmetry.Ward.identity}) is implied by the
Ehrenfest theorem for the equations of motion.
The vacuum-to-vacuum scattering amplitude is~\footnote{Having in mind the invariant path integral measure for fermions given in footnote~6, the
generalisation for fermions of the derivation leading to the identity~(\ref{fundamental.Ward.identity})
is straightforward and we do not consider it here separately;
see, however, section~\ref{Conformal anomaly in Yukawa theory} below.},
\begin{equation}
\label{pathintegral.lagrangian}
\langle {\it in}|{\it out}\rangle = \int \bar{\mathcal{D}}\phi\, e^{iS_\phi}\,,
\end{equation}
where $\bar{\mathcal{D}}\phi$ is the Weyl invariant measure given in Eq.~(\ref{invariant measure}) and 
$S_\phi$ is a conformal scalar action, whose kinetic part is given by~(\ref{Scalar.Action.1}).

Requiring that infinitesimal Weyl transformations, $\Omega(x) \rightarrow 1 + \omega(x)$, under which the fields transform as,
\begin{equation}
  \phi\rightarrow \phi^\prime = \phi -\frac{D-2}{2}\omega \phi
\,,\quad
  g_{\mu\nu}\rightarrow g_{\mu\nu}^\prime = g_{\mu\nu} +2\omega g_{\mu\nu}
\,,\quad
 \Gamma^\alpha_{\;\mu\nu}\rightarrow
     {\Gamma^\prime}^\alpha_{\;\mu\nu} = \Gamma^\alpha_{\;\mu\nu} +\delta^\alpha_{\;\mu}\partial_\nu \omega
\,,
\label{infinitesimal Weyl transformations}
\end{equation}
do not change the {\it in-out} amplitude~(\ref{pathintegral.lagrangian}) yields,
\begin{eqnarray}
\label{pathintegral.ward.1}
\langle {\it in}|{\it out}\rangle
&=& \int \bar{\mathcal{D}}\phi' e^{iS^\prime_\phi}
% =  \int \mathcal{D}\varphi e^{i\int \text{d}^Dx \mathcal{L'}} =
\\
&=&\int \bar{\mathcal{D}}\phi e^{i S_\phi}
\Bigg [1 + i\int\text{d}^Dx\sqrt{-g}\Bigg (\!-\frac{D-2}{2\sqrt{-g}}\frac{\delta S_\phi}{\delta \phi(x)} \omega(x)\phi(x)
+ \frac{2}{\sqrt{-g}}\frac{\delta S_\phi}{\delta g^{\mu\nu}(x)} \omega(x)g^{\mu\nu} \nonumber\\
&&\hskip 2cm
+ \bar{\nabla}_\mu \left (\frac{1}{\sqrt{-g}}\frac{\delta S_\phi}{\delta \mathcal{T}_{\mu}(x)}\right ) \omega(x)\Bigg )\Bigg ]
\nonumber
\,.
\end{eqnarray}
Since this must be true for any arbitrary infinitesimal $\omega(x)$, Eq.~(\ref{pathintegral.ward.1}) then implies,
\begin{equation}
\label{Ward.Identities}
 \int \bar{\mathcal{D}}\phi e^{i S_\phi}  \left (T_\mu^\mu + \bar{\nabla}_\mu \Pi^\mu\right ) = 0\,,
\end{equation}
where,
\begin{eqnarray}
\label{energy.momentum.definition}
T_{\mu\nu} = \frac{2}{\sqrt{-g}} \frac{\delta S}{\delta g^{\mu\nu}}\,,\\
\label{torsion.source.definition}
\Pi^\mu = \frac{1}{\sqrt{-g}}\frac{\delta S}{\delta \mathcal{T}_{\mu}(x)}\,,
\end{eqnarray}
and we have used the Ehrenfest theorem~(\ref{Lemma.Two.Points.Functions}) from Appendix~B.
Upon dividing~(\ref{Ward.Identities}) by $\langle {\it in}|{\it out}\rangle$ we finally get,
\begin{equation}
 \langle T_\mu^\mu\rangle  +\langle \bar{\nabla}_\mu \Pi^\mu\rangle = 0
 \,,
 \label{fundamental.Ward.identity}
\end{equation}
proving thus~(\ref{conformal.symmetry.Ward.identity}).
The angular brackets in~(\ref{fundamental.Ward.identity}) denote an expectation value of the time-ordered product
and all the derivatives must be evaluated inside the time-ordered product.
The identity~(\ref{fundamental.Ward.identity}) is the main result of this work.
In order to elucidate its meaning, in the remainder of this section
we discuss some useful examples.

\subsection{Conformal anomaly in an interacting scalar theory}
\label{Conformal anomaly in an interacting scalar theory}

Consider now the following self-interacting scalar theory,
\begin{equation}
\label{dynamical.scalar.theory}
S_\phi = \int \text{d}^D x \sqrt{-g} \Big(\frac{1}{2}g^{\mu\nu} \bar{\nabla}{}_\mu\phi \bar{\nabla}{}_\nu\phi - \lambda \phi^4\Big)
\,,
\end{equation}
which is conformal in $D=4$.
The action~(\ref{dynamical.scalar.theory}) then implies,
\begin{eqnarray}
\label{energy.momentum.source}
  T_\mu^\mu \!&=&\! -\bar{\nabla}{}_\mu\phi \bar{\nabla}{}^\mu\phi + D \lambda \phi^4\,,\\
\label{Torsion.source.scalar.dynamical}
  \bar{\nabla}_\mu \Pi^\mu \!&=&\! \bar{\nabla}{}_\mu\phi \bar{\nabla}{}^\mu\phi +\phi \bar{\nabla}{}_\mu\bar{\nabla}{}^\mu\phi \,,
\end{eqnarray}
and the identity~(\ref{conformal.symmetry.Ward.identity}) leads to,
\begin{equation}
\label{scalar.fundamental.identity}
\big\langle\phi \left (\bar{\nabla}{}_\mu\bar{\nabla}{}^\mu\phi + 4 \lambda \phi^3\right )\rangle
+ (D-4)  \lambda \big\langle \phi^4\big\rangle
= (D-4) \lambda  \big\langle\phi^4\big\rangle\,,
\end{equation}
where we applied again Eq.~(\ref{Lemma.Two.Points.Functions}). Na\"ively one might think that the term
on the right hand side may generate a finite contribution
in dimensional regularisation,
as $\lambda \langle \phi^4\rangle \propto 1/(D-4)$. However,
this cannot be so since $ \lambda \langle\phi^4\rangle$ contributes to the energy-momentum
tensor as, $\langle T_{\mu\nu}\rangle  \supset g_{\mu\nu}\lambda\langle  \phi^4\rangle$,
such that any primitive divergence must be regulated, implying that $ \lambda \langle\phi^4\rangle$ must be finite.
This immediately implies that, after regularisation and when the limit $D\rightarrow 4$ is taken, the term
$(D-4) \lambda\langle  \phi^4\rangle$ in~(\ref{scalar.fundamental.identity}) will vanish
and thus the fundamental Ward identity~(\ref{conformal.symmetry.Ward.identity}) will be respected.

\subsection{Conformal anomaly in Yukawa theory}
\label{Conformal anomaly in Yukawa theory}

In order to further motivate the identity~(\ref{conformal.symmetry.Ward.identity}), let us consider the following Yukawa theory, whose action is conformal in $D=4$,
\begin{equation}
\label{Conformal.Yukawa}
S_{\rm Yu} = \int \text{d}^D x \sqrt{-g} \left ( \frac{1}{2}\bar{\nabla}{}_\mu\phi\bar{\nabla}{}^\mu\phi + \frac{i}{2}( \bar{\psi} \gamma^\mu \overset{\leftrightarrow}{\nabla}{}_\mu\psi) - y\phi \bar{\psi}\psi\right )
\,,
\end{equation}
where $\psi$ and $\phi$ represent fermionic and scalar fields and $y$ is a (constant) Yukawa coupling.
For this theory the energy-momentum tensor and divergence of the torsion source are given by,
\begin{eqnarray}
\label{energy.momentum.tensor.Yukawa}
T_{\mu\nu}&=& \frac{i}{2} \left ( \bar{\psi}\gamma_{(\mu}\overset{\leftrightarrow}{\nabla}{}_{\nu)} \psi \right ) + \bar{\nabla}{}_\mu\phi \bar{\nabla}{}_\nu\phi - g_{\mu\nu} \left (\frac{1}{2}\bar{\nabla}{}_\alpha\phi\bar{\nabla}{}^\alpha\phi + \frac{i}{2}( \bar{\psi} \gamma^\alpha \overset{\leftrightarrow}{\nabla}{}_\alpha\psi) - y\phi \bar{\psi}\psi\right )
\,,\\
\label{Torsion.Source.Yukawa}\bar{\nabla}{}_\mu\Pi^\mu &=& \bar{\nabla}{}_\mu\phi \bar{\nabla}{}^\mu\phi + \phi \bar{\nabla}{}_\mu\bar{\nabla}{}^\mu\phi
\,.
\end{eqnarray}
Summing~(\ref{energy.momentum.tensor.Yukawa}) and~(\ref{Torsion.Source.Yukawa}) then leads to,
\begin{equation}
\label{proof.conformality.yukawa}
 -\frac{3  }{2} \Big\langle\bar{\psi}\left ( i \cancel{\nabla} \psi - y\phi \psi\right )\Big\rangle
+ \frac{3  }{2} \Big\langle\big( i \bar{\psi}\overset{\leftarrow}{\cancel{\nabla}}+y\phi \bar{\psi} \big)\psi\Big\rangle
+ \Big\langle\phi\left (  \bar{\nabla}{}_\mu\bar{\nabla}{}^\mu\phi +y\bar{\psi}\psi\right )\Big\rangle
 + (D-4) y\big\langle\phi \bar{\psi}\psi\big\rangle
\,.
\end{equation}
The first three angular brackets in~(\ref{proof.conformality.yukawa})
constitute expectation values of composite operators, each containing a product of a field
and an equation of motion for either $\psi,\,\bar{\psi}$ or $\phi$. Therefore, all of them must vanish by
(the Yukawa-theory version of) the Ehrenfest theorem, {\it c.f.} Eq.~(\ref{Lemma.Two.Points.Functions}).
Similarly as above one can argue that the term $(D-4) y\langle\phi \bar{\psi}\psi\rangle$ in~(\ref{proof.conformality.yukawa})
must vanish in $D=4$. Indeed, if that was not the case it would have lead to a nonvanishing
divergence in the energy momentum tensor which contains
a term of the form, $\langle T_{\mu\nu}\rangle \supset g_{\mu\nu} y\langle\phi \bar{\psi}\psi\rangle$.

\subsection{Conformal anomaly in an interacting Yang-Mills}
\label{Conformal anomaly in an interacting non-Abeliean gauge theory}

Another type of interacting theory that occurs in the standard model is a gauge theory that couples 
to a charged scalar current.~\footnote{The coupling to fermions can be also included and we leave it to the reader as 
an exercise. One can in fact consider the action~(\ref{Conformal.Yukawa}) with minimal coupling to gauge fields, $\nabla_\mu \rightarrow D_\mu$. 
Proceeding in the analogous way as we do here, one would then find an extra contribution to Eq.~(\ref{fundamental ward identity:YM}) proportional to 
$(D-4) \langle A_\mu \bar{\psi}\gamma^\mu\psi\rangle$, which again drops in $D=4$ since the operator $ \langle A_\mu \bar{\psi}\gamma^\mu\psi\rangle$ must be finite.} 
The action is,
\begin{equation}
 S_{YM} = \int d^Dx\sqrt{-g}{\rm Tr}\left(-\frac14 g^{\mu\rho}
                       g^{\nu\sigma}F_{\mu\nu}F_{\rho\sigma} 
  + g^{\mu\nu}\big(\bar{D}_\mu\phi\big)^\dagger\big(\bar{D}_\nu\phi\big)\right)
\label{action YM}
\end{equation}
where the gauge field is taken to be in adjoint representation and the scalar in fundamental representation
(just like as they are in the standard model), and the $\rm Tr$ is taken on the group indices.
The field strength is given by, 
\begin{equation}
F_{\mu\nu}= \bar\nabla_\mu A_\nu - \bar\nabla_\nu A_\mu + e\big [A_\mu, A_\nu\big ] = F_{\mu\nu}^a \lambda^a_{\rm adj}
\,,\quad
F_{\mu\nu}^a = \bar\nabla_\mu A_\nu^a - \bar\nabla_\nu A_\mu^a +ef^{abc}A_\mu^b A_\nu^c
\,,
\label{field strength}
\end{equation}
where $f^{abc}$ are the adjoint representation structure constants, $\lambda^a_{\rm adj}$ the group generators also in the adjoint representation,
$e$ is the gauge coupling constant and 
\begin{equation}
\bar\nabla_{[\mu} A_{\nu]}^a = \partial_{[\mu} A_{\nu]}^a +\frac{D\!-\!4}{2}\mathcal{T}_{[\mu} A_{\nu]}^a
\label{conformal gauge derivative}
\end{equation}
is the conformal exterior derivative as it acts on the gauge field. 
Note that the conformal derivative become the usual exterior derivative in $D=4$, but 
it breaks gauge symmetry away from $D=4$.
Since we are ultimately interested in $D=4$, the conformal derivative for gauge fields 
was introduced for regularisation purposes only.
Varying the action~(\ref{action YM}) with respect to the matter fields $A_\mu$, $\phi$ 
(here $\phi=\phi^a \lambda^a_{\rm fund}$, $\lambda^a_{\rm fund}$ are the basis matrices 
of the fundamental representation of the gauge group) 
and $\phi^\dagger$  gives the following equations of motion,
\begin{eqnarray}
  \bar D_{\mu}\left (\bar D^\mu \phi\right ) \!&=&\!   0
,\quad 
  \bar D_{\mu}\left (\bar D^\mu \phi\right )^\dagger  = 0\,,
\label{scalar EOM}
\\
\quad \bar D_{\mu} &\!=\!& \overset{\circ}{\nabla}_\mu \!-\!\frac{D\!-\!2}{2}\mathcal{T}_\mu \!+\! i e A_\mu
,\quad \bar D^{\mu} = g^{\mu\nu}\big(\overset{\circ}{\nabla}_\nu \!+\! \frac{D\!-\!2}{2}\mathcal{T}_\nu \!+\! i e A_\nu\big)
\nonumber\\
\bar D_\mu F^{\mu\nu}   
   \!&=&\! ie g^{\mu\nu}\Big[\phi^\dagger \bar D_\nu\phi-\big(\bar D_\nu\phi\big)^\dagger \phi\Big]
\,,
\label{gauge field EOM}
\end{eqnarray}
where $\overset{\circ}{\nabla}_\mu$ is the general relativity covariant derivative, that is the space-time covariant derivative
computed using the Christoffel symbols. When acting on a scalar, it equals the partial derivative $\overset{\circ}{\nabla}_\mu \phi = \partial_\mu \phi$, while 
when acting on a vector it has the expression, $\overset{\circ}{\nabla}_\mu V^\mu = \frac{1}{\sqrt{-g}}\partial_\mu \left (\sqrt{-g} V^\mu\right )$.
Note that in~(\ref{scalar EOM}--\ref{gauge field EOM})
the conformal covariant derivative always comes with conformal weight of the associated tensor density.
Thus conformal weight of $\phi$ is $-(D-2)/2$, of $\sqrt{-g}\bar D^\mu\phi$ is $+(D-2)/2$ while conformal weight of 
$\sqrt{-g}F^{\mu\nu}$ is $+(D-4)/2$. 
On the other hand, varying~(\ref{action YM}) with respect to the geometric fields $g^{\mu\nu}$ and $\mathcal{T}_\mu$ results in,
\begin{eqnarray}
 T_{\mu\nu}\!\!&=&\!\!{\rm Tr}\Bigg\{ \!F_{\mu\alpha}F_{\nu\beta} g^{\alpha\beta}
           \!-\! \frac14 g_{\mu\nu} F_{\alpha\beta}F_{\gamma\delta}  g^{\alpha\gamma}  g^{\beta\delta}
          \!+\! 2\big(\bar D_{\mu}\phi\big)^\dagger\big(\bar D_{\nu}\phi\big)
                     \!-\!g_{\mu\nu}\Big[g^{\alpha\beta}\big(\bar D_{\alpha}\phi\big)^\dagger\big(\bar D_{\beta}\phi\big)\Big]\!\bigg\}
\label{Tmn:YM}
\\
\Pi^\mu \!&=&\!{\rm Tr}\Bigg\{  -\frac{D\!-\!4}{2} A_\nu F_{\gamma\delta} g^{\nu\delta}g^{\mu\gamma}
  +\frac{D\!-\!2}{2}g^{\mu\nu}\Big[\phi^\dagger\big(\bar D_{\nu}\phi\big)
                                                         +\big(\bar D_{\nu}\phi\big)^\dagger\phi\Big]\Bigg\}
\,.
\label{Pim:YM}
\end{eqnarray}
Taking a trace and expectation value~\footnote{In taking an expectation value one ought to integrate over 
fields fluctuations weight by the action. By making use of the phase space version of the path integral quantisation,
 in Appendix~A we show that such a quantisation respects conformal symmetry  in Abelian gauge 
theories in the sense that the corresponding path integral measure is conformal. 
Here we assume that that is also the case with the measure 
of non-Abelian gauge fields, but leave the proof to future publication.}
  of $T_{\mu\nu}$ we get,
\begin{eqnarray}
     \langle T^\mu_\mu\rangle ={\rm Tr}\Bigg\{\frac{D-4}4 \langle F_{\mu\nu}F_{\rho\sigma}\rangle g^{\mu\rho}g^{\nu\sigma}
           -(D\!-\!2)\langle\big(\bar D_{\mu}\phi\big)^\dagger\big(\bar D_{\nu}\phi\big)\rangle\Bigg\}
\label{Tmm:YM}
\end{eqnarray}
while taking a (conformal)~\footnote{The conformal weight of $\sqrt{-g}\Pi^\mu$ is zero, such that 
taking covariant and conformal covariant derivatives of $\Pi^\mu$ coincide.}
 covariant divergence and expectation value of $\Pi^\mu$ we obtain,
\begin{eqnarray}
      \langle \bar\nabla_\mu \Pi^\mu\rangle
  \!&=&\!{\rm Tr}\Bigg\{-\frac{D-4}4  \langle F_{\mu\nu}F_{\rho\sigma}\rangle g^{\mu\rho}g^{\nu\sigma}
   + (D\!-\!2)\langle\big(\bar D_{\mu}\phi\big)^\dagger\big(\bar D_{\nu}\phi\big)\rangle
\nonumber\\
 && -\frac{D-4}2  i e\Big\langle A_\mu\Big[\phi^\dagger \big(\bar D^{\mu}\phi\big)
                                               - \big(\bar D^{\mu}\phi)^\dagger \phi\Big]\Big\rangle\Bigg\}
\,.
\label{nabla Pim:YM}
\end{eqnarray}
By combining~(\ref{Tmn:YM}) and~(\ref{Pim:YM}) we see that the fundamental identity~(\ref{conformal.symmetry.Ward.identity}) is not quite satisfied,
\begin{eqnarray}
      \langle \bar\nabla_\mu \Pi^\mu\rangle +  \langle T^\mu_\mu\rangle
  = -\frac{D-4}2 i e{\rm Tr}\Big\langle A_\mu\Big[\phi^\dagger \big(\bar D^{\mu}\phi\big)
                                           - \big(\bar D^{\mu}\phi)^\dagger \phi\Big]\Big\rangle+\frac{D-4}2 e \langle A_\mu \bar{\psi}\gamma^\mu\psi\rangle
\,.
\label{fundamental ward identity:YM}
\end{eqnarray}
To see that the terms on the right hand side must vanish, note that the last term in 
the (expectation value of the) energy-momentum tensor~(\ref{Tmn:YM}) can be decomposed as, 
\begin{equation}
{\rm Tr}\Big\langle \big(\bar D_{\mu}\phi\big)^\dagger\big(\bar D^{\mu}\phi\big)\Big\rangle
    = {\rm Tr}\Big\langle \big(\bar \nabla_{\mu}\phi\big)^\dagger\big(\bar \nabla^{\mu}\phi\big)\Big\rangle
- ie{\rm Tr}\Big\langle A_\mu\big[\phi^\dagger\big(\bar D^{\mu}\phi\big)-\big(\bar D^{\mu}\phi\big)^\dagger\phi\big]\Big\rangle
\,.
\label{Tmn:part}
\end{equation}
Upon renormalisation is exacted both terms must be finite. To see that observe that 
the first term in~(\ref{Tmn:part}) is present also when the gauge field vanishes and therefore must be finite by itself 
and thus the second term must be also finite (because the whole energy-momentum tensor must be finite),
 implying that the right hand side of~(\ref{fundamental ward identity:YM}) must vanish
and therefore the fundamental conformal Ward identity~(\ref{conformal.symmetry.Ward.identity}) is satisfied. 
Analogously, since $\langle A_\mu \bar{\psi}\gamma^\mu\psi\rangle$ also appears in the energy momentum tensor, 
with a contribution that vanishes upon setting $A_\mu$ to zero, and so it must be finite by itself to guarantee that the energy momentum 
tensor is finite.

In conclusion, we have considered {\it all} types of 
 interacting field theories that occur in the standard model and we have showed that
 the fundamental identity~(\ref{conformal.symmetry.Ward.identity}) is satisfied for all of them, implying that it is
 also satisfied in the standard model.

\subsection{Boundary terms and local anomaly}
\label{Boundary (total derivative) terms}

There are terms that contribute to $\langle T^\mu_{\;\mu}\rangle$ as total derivatives.
To expound on the meaning of such terms, let us consider
the scalar 1-loop effective action around a general gravitational background which
before regularisation is of the form~\cite{Shapiro:2001rz},
\begin{equation}
\label{1-loop-effective-action}
\Gamma_{\phi}=\frac{1}{4\pi^2} \int \text{d}^D x\sqrt{-g}
\left [ \Gamma\left (1\!-\!\frac{D}{2}\right ) \left (\alpha\bar{R}\right )^{\frac{D}{2}}
 \!+\! \Gamma\left (2\!-\!\frac{D}{2}\right )\left ( \beta C_{\alpha\beta\gamma\delta}C^{\alpha\beta\gamma\delta}
 \!+\! \gamma\mathcal{E}_4\right )\left (\alpha\bar{R}\right )^{\frac{D-4}{2}}\right ],
\end{equation}
where $\bar{R}$ is the Ricci scalar formed from the curvature tensor with torsion,
$C_{\alpha\beta\gamma\delta}$ is the Weyl tensor which is independent of the torsion trace,
\begin{equation}
\label{Euler.density}
\mathcal{E}_4 = \frac{1}{4!}\epsilon^{\mu\nu\lambda\sigma}\epsilon_{\alpha\beta}{}^{\gamma\delta}
                  \bar{R}{}^\alpha{}_{\gamma\mu\nu} \bar{R}{}^\beta{}_{\delta\lambda\sigma}
                  \implies
\frac{1}{D!}\epsilon^{\mu\nu\lambda\sigma\rho_1\cdots\rho_{D-4}}
 \epsilon_{\rho_1\cdots\rho_{D-4}\alpha\beta}{}^{\gamma\delta}
 \bar{R}{}^\alpha{}_{\gamma\mu\nu} \bar{R}{}^\beta{}_{\delta\lambda\sigma}  \,,
\end{equation}
is the Euler density which is in four dimensions a total divergence~\cite{Nieh:1979hf}
and $\alpha,\beta$ and $\gamma$ are constants.
The action~(\ref{1-loop-effective-action}) is divergent (in the sense that it  yields divergent contributions
to the Einstein's equation) and thus it ought to be renormalised.
The first step in the renormalisation procedure is to identify the finite parts of the action.

To do that let us firstly analyse the contribution to the stress-energy tensor from the Euler density~(\ref{Euler.density}).
Its variation gives a finite contribution to the stress-energy tensor and as such does not need any counter term.
To see that let us vary the contribution of $\mathcal{E}_4$ to the effective action~(\ref{1-loop-effective-action}). We have,
\begin{eqnarray}
&&\hskip 0cm
\frac{\delta}{\delta g^{\rho\tau}(z)} \int \text{d}^D x\sqrt{-g}\epsilon^{\mu\nu\lambda\sigma\rho_1\cdots\rho_{D-4}}\epsilon_{\rho_1\cdots\rho_{D-4}\alpha\beta}{}^{\gamma\delta} \bar{R}^\alpha{}_{\gamma\mu\nu}\bar{R}^\beta{}_{\delta\lambda\sigma}\\
\label{energy.density.of.euler.density}
&&\hskip 6cm
=\int \text{d}^D x\sqrt{-g}\epsilon^{\mu\nu\lambda\sigma\rho_1\cdots\rho_{D-4}}\frac{\delta\left (\epsilon_{\rho_1\cdots\rho_{D-4}\alpha\beta}{}^{\gamma\delta}\right)}{\delta g^{\rho\tau}(z)} \bar{R}^\alpha{}_{\gamma\mu\nu}\bar{R}^\beta{}_{\delta\lambda\sigma}
\,,
\nonumber
\end{eqnarray}
were we dropped the following two terms,
\[\frac{\delta\left (\sqrt{-g}\epsilon^{\mu\nu\lambda\sigma\rho_1\cdots\rho_{D-4}}\right)}{\delta g^{\rho\tau}(z)}
\hskip0.5cm \&\hskip0.5cm\sqrt{-g}\epsilon^{\mu\nu\lambda\sigma\rho_1\cdots\rho_{D-4}}  \frac{\delta\left ( \bar{R}{}^\alpha{}_{\gamma\mu\nu}\right)}{\delta g^{\rho\tau}(z)} \bar{R}{}^\beta{}_{\delta\lambda\sigma}\,,\]
the first one because the factors of $\sqrt{-g}$ cancel between $\sqrt{-g}$ and
the Levi-Civit\`a tensor and the second one because it vanishes due to the Bianchi identities.
The term that is left in~(\ref{energy.density.of.euler.density})
%$\frac{\delta\left (\epsilon_{\rho_1\cdots\rho_{D-4}\alpha\beta}{}^{\gamma\delta}\right)}{\delta g^{\mu\nu}(z)}$
is identically zero in $D=4$ since in four dimensions $\epsilon_{\alpha\beta}{}^{\gamma\delta}$ yields contributions
that are independent of the metric tensor. Taking account of this we finally arrive at the expression,
\begin{equation}
\begin{split}
\label{energy.momentum.tensor.of.euler.density}
&\hskip -0cm
\frac{2}{\sqrt{-g}}\frac{\delta}{\delta g^{\rho\tau}(z)} \int \text{d}^D x\sqrt{-g}\epsilon^{\mu\nu\lambda\sigma\rho_1\cdots\rho_{D-4}}\epsilon_{\rho_1\cdots\rho_{D-4}\alpha\beta}{}^{\gamma\delta}
\bar R^\alpha{}_{\gamma\mu\nu}\bar R^\beta{}_{\delta\lambda\sigma}\\
\nonumber
&\hskip 6cm=\frac{D-4}{D} g_{\rho\tau}  \left (\epsilon^{\mu\nu\lambda\sigma}\epsilon_{\alpha\beta}{}^{\gamma\delta}
\bar R^\alpha{}_{\gamma\mu\nu}\bar R^\beta{}_{\delta\lambda\sigma}\right )
=\frac{D-4}{D} g_{\rho\tau}\mathcal{E}_4
\,,
\end{split}
\end{equation}
which can be verified directly from~(\ref{energy.density.of.euler.density}) by evaluating $\delta \big(\epsilon_{\rho_1\cdots\rho_{D-4}\alpha\beta}{}^{\gamma\delta}\big)/\delta g^{\rho\tau}(z)$. This shows that we get a finite contribution to the stress-energy tensor from the divergent contribution proportional to the Euler density term in the effective action~(\ref{1-loop-effective-action}) and thus we do not have to add a counter term to renormalise it.

To be consistent, we should also check that the same term gives a finite contribution to the Weyl field source, $\Pi^\mu$. 
Indeed, upon noticing that ${\cal E}_4 = \bar{\nabla}_\mu \mathcal{V}^{\mu}$, where 
${\cal V}^\mu$ has scaling dimension $-4$ under Weyl transformations, we can see that this is the case. 
Using the conformal Stokes theorem~(\ref{stokes.theorem.2}) one can show that in general $D$, 
\begin{equation}
\label{dimension.Euler.Vector} \bar{\nabla}_\mu \mathcal{V}^{\mu} = \frac{1}{\sqrt{-g}}\partial_\mu \left (\sqrt{-g}  \mathcal{V}^{\mu} \right ) -(D-4) T_\mu \mathcal{V}^\mu\,, 
\end{equation}
since the length dimension of $\mathcal{V}^\mu$ is $-3$~(it contains 3 derivatives acting on the metric), and thus $\int\text{d}^Dx \sqrt{-g} \bar{\nabla}_\mu \mathcal{V}^{\mu}$ is 
only dimensionless in $D=4$. We can then conclude that, since the first term in Eq.~(\ref{dimension.Euler.Vector}) 
is a boundary term in any dimension, the Euler density contribution to the torsion source is,
\begin{equation}
\label{dilaton.source.euler.density} \frac{1}{\sqrt{-g}}\frac{\delta}{\delta \mathcal{T}_\mu} \int \text{d}^D x\sqrt{-g} \mathcal{E}_4 =  -(D-4) \mathcal{V}{}^\mu
\,,
\end{equation}
which shows that the fundamental Ward identity~(\ref{conformal.symmetry.Ward.identity})
 is in fact respected by this contribution. 
Note that this is not possible to achieve in a theory containing the metric only, since necessarily the Gauss-Bonnet
contribution is finite and spoils the identity $\langle T_\mu^\mu\rangle =0$. 

However, this is not yet the end of the story. In order to renormalise~(\ref{1-loop-effective-action}) one has to make
all the terms (except possibly ${\mathcal E}_4)$ finite in $D=4$ and the only way of doing that within 
dimensional regularisation
is to add scale dependent counterterms that in this way introduce a scale dependence in the renormalized
action, thus breaking conformal symmetry. In the next section we discuss how to cure such an apparent violation of 
conformal Ward identities.

\subsection{The role of scale $\mu$}
\label{The role of scale mu}

In subsection~\ref{Boundary (total derivative) terms} we have seen that -- in presence of
geometric or scalar field condensates,  perturbative renormalised
effective actions typically contain scale dependent terms that may violate the
conformal Ward identity~(\ref{conformal.symmetry.Ward.identity}). 
Indeed, consider the first term in the effective action~(\ref{1-loop-effective-action}).
Due to the presence of $\Gamma(1\!-\!D/2)\propto 1/(D\!-\!4)$, that term is divergent in $D=4$,
Using standard dimensional regularisation, by which one adds a scale dependent countertem $\propto1/(D\!-\!4)$
to remove the divergence, one ends up with the following renormalised, scale dependent, effective action,
\begin{equation}
 \Gamma_{\rm eff} =\int d^4 x\sqrt{-g} {\cal L}_{\rm eff} 
 =  \int d^4 x\sqrt{-g} \tilde \alpha \bar R^2\ln\bigg(\frac{\bar R}{\mu^2}\bigg)
\,,
\label{1 loop renormalized action:b}
\end{equation}
where $\tilde \alpha$ is a (finite) coupling constant that does not depend on $\mu$. 
Upon varying this action with respect to $g_{\mu\nu}$ and taking a trace, then upon varying with respect to ${\cal T}_\mu$
 and taking a divergence, one ends up with, 
\begin{equation}
\label{Ward.identity.satisfied}
g^{\mu\nu} \frac{2}{\sqrt{-g}} \frac{\delta \Gamma_{\rm eff}}{\delta g^{\mu\nu}} 
 + \bar{\nabla}_\alpha \frac{1}{\sqrt{-g}} \frac{\delta \Gamma_{\rm eff}}{\delta \mathcal{T}_\alpha} 
= 2\tilde\alpha \bar{R}^2
 = - \mu \frac{\text{d}\mathcal{L}_{\rm eff}}{\text{d}\mu}
\,.
\end{equation}
This shows that a standard renormalisation procedure (such as dimensional regularisation) leads to 
scale-dependent effective action that breaks 
the conformal Ward identity~(\ref{Ward.identity.satisfied}), see Eq.~(\ref{conformal.symmetry.Ward.identity}).
In order to understand  how to deal with such a situation without breaking the local Weyl symmetry, we should 
carefully ponder on the meaning of the renormalisation scale $\mu$. 

From the perspective of renormalisation, $\mu$ is not a physical quantity, and it can be chosen 
to correspond to {\it any} scale we pick to probe the physics. 
On the other hand, renormalisation group flow is constructed 
by demanding that changing this scale does not change the physics, which is expressed by the following 
requirement on an effective action $\Gamma$,
\begin{equation}
\mu\frac{\text{d}{\Gamma[\mu]}}{\text{d}\mu} = 0
\,.
\label{scale independence of Gamma}
\end{equation}
This means that the effective action $\Gamma[\mu]$, and consequently all observables derived from it, cannot 
dependent on $\mu$.~\footnote{ 
This statement should not be mixed with the statement that {\it e.g.} 
coupling constants (which determine physical scattering rates) run 
with some physical scale. When such statements are made, it is always assumed that the scale has a physical 
meaning, an example being an invariant energy in scattering processes. In fact, one should 
understand~(\ref{scale independence of Gamma}) as follows. There exists a map that 
generates a characteristic flow of the effective action which 
maps $\Gamma[\mu,\lambda_i,\psi_j]$ onto $\Gamma[\mu^\prime,\lambda_i^\prime,\psi_j^\prime]$ such that 
 $\Gamma[\mu,\lambda_i,\psi_j]=\Gamma[\mu^\prime,\lambda_i^\prime,\psi_j^\prime]$.
Since in general the couplings $\lambda_i\neq \lambda_i^\prime$ and the fields $\psi_j\neq\psi_j^\prime$,
the form ({\it i.e.} the dependence on the couplings and fields) of $\Gamma[\mu,\lambda_i,\psi_j]$ 
and $\Gamma[\mu^\prime,\lambda_i^\prime,\psi_j^\prime]$ 
are not in general the same. Nevertheless, both actions are equivalent in
the sense that they describe exactly the same physics. 
} 

As a consequence, we typically require $\mu$ to be a spacetime constant, $\partial_\alpha \mu=0$.
That is just a probing mass scale which does not change the dynamics. 
In this context of this work however, $\partial_\alpha \mu =0$ is not a gauge invariant statement, 
in the sense that it transforms non-trivially under Weyl transformations.
Therefore, it is much more natural to demand instead,
\begin{equation}
  \bar{\nabla}_\alpha \mu = \partial_\alpha\mu +\frac{D-2}{2} {\cal T}_\alpha \mu   = 0
\,, 
\label{constancy of mu}
\end{equation}
which is the conformally invariant expression for spacetime constancy of $\mu$. 
In writing~(\ref{constancy of mu}) we have assumed that the conformal dimension of $\mu$ is equal to that of 
a classical scalar field, {\it i.e.}  $w_\mu=-(D-2)/2$, which is the natural choice. 
When Eq.~(\ref{constancy of mu}) is enforced on $\mu$ we have (on-shell), 
\begin{equation}
  \mathcal{T}_\alpha = -\frac{2}{D-2} \partial_\alpha \log(\mu)
\,. 
\label{constancy of mu:2}
\end{equation}
Note that $\mathcal{T}_\alpha $ is purely longitudinal, {\it i.e.} its transverse part is zero.
Eq.~(\ref{constancy of mu:2}) does not really mean that $\mu$ is spacetime dependent. 
Indeed all that dependence is pure gauge 
as there always exists a (gauge) frame defined by $\mathcal{T}_\alpha = 0$ in which $\mu$ is truly constant. 
The true spacetime constancy of $\mu$ is therefore retained by the gauge invariant statetement, 
$\frac{1}{\mu}\bar{\nabla}_\alpha \mu = 0$ and any other choice of $\mu$ (including the standard one) 
will be gauge dependent and thus will break conformal symmetry.  
The problem with the standard choice $\mu={\rm constant}$ in the effective action is that 
off-shell $\mu=\mu[\mathcal{T}_\mu]$ and neglecting that dependence would be simply wrong
(in the sense that it breaks conformal symmetry). In other words, the gauge $\mathcal{T}_\mu=0$ 
is allowed only after variation is exacted, {\it i.e.} {\it on-shell}.

That~(\ref{constancy of mu}) is the natural choice can be argued as follows.
Since at low energies the Weyl symmetry is (spontaneously) broken, there 
exists a Goldstone boson $\theta$ which transforms under the Weyl scaling as, 
$\theta \rightarrow \theta + \log \Omega$. 
Then the quantity $e^{\frac{D-2}{2}\theta}\mu$ is a (gauge-invariant) scalar field combination
 and therefore can be used as an invariant probe of some physical scale.

As a consequence of the requirement~(\ref{constancy of mu}), the torsion source $\Pi^\mu$
acquires an extra contribution $\Delta_\mu\Pi^\alpha$ of the form,~\footnote{
Note that in Eq.~(\ref{Non-Local.Torsion.Source}) the derivative operator and inverse Laplacian 
are torsion independent and contain only contributions on the metric. However, since the scaling dimension of 
$\mu\frac{\partial \mathcal{L}_{\rm eff}}{\partial \mu}$ is equal to $-D$, the conformal divergence of $\Delta_\mu\Pi^\alpha$
satisfies $\overset{\circ}{\nabla}_\alpha \Delta_\mu\Pi^\alpha =\bar{\nabla}_\alpha \Delta_\mu\Pi^\alpha = \mu\frac{\partial \mathcal{L}_{\rm eff}}{\partial \mu} $, as in Eq.~(\ref{Non-Local.Torsion.Source:2}). 
This in turn implies that $\Delta_\mu\Pi^\alpha$ scales as $\mu\frac{\partial \mathcal{L}_{\rm eff}}{\partial \mu}$, 
that is with conformal weight $-D$, even though this is not manifest in Eq.~(\ref{Non-Local.Torsion.Source}). 
}
\begin{equation}
\label{Non-Local.Torsion.Source} 
\Delta_\mu\Pi^\alpha =-\frac{2}{D-2}\overset{\circ}{\nabla}{}^\alpha\frac{1}{{\overset{\circ}{\Box}}} \mu 
           \frac{\partial \mathcal{L}_{\rm eff}}{\partial \mu}
 \,.
\end{equation} 
Taking a (conformal) divergence of $\Delta_\mu\Pi^\alpha$ yields, 
\begin{equation}
\label{Non-Local.Torsion.Source:2} 
\bar\nabla_\alpha\big(\Delta_\mu\Pi^\alpha\big)=\overset{\circ}{\nabla}_\alpha\big(\Delta_\mu\Pi^\alpha\big) =-\frac{2}{D-2}\mu 
           \frac{\partial \mathcal{L}_{\rm eff}}{\partial \mu}
 \,,
\end{equation} 
where we took account of the fact that the conformal weight of torsion source $\sqrt{-g}\Pi^\alpha$ is zero.

Adding~(\ref{Non-Local.Torsion.Source:2}) to~(\ref{Ward.identity.satisfied})
and taking $D=4$
we obtain that the fundamental Ward identity~(\ref{conformal.symmetry.Ward.identity})
is satisfied. 
This is not a coincidence and in fact it works in general. Namely, demanding that regularisation respects 
conformal symmetry by requiring that any scale dependence introduced by regularisation procedure 
is conformally invariant will result in an effective action which satisfies 
the fundamental conformal Ward identity~(\ref{conformal.symmetry.Ward.identity}).

In the above procedure we have introduced non-locality in the source of torsion 
such that we do no longer meet the
requirement that $T_\mu^\mu = -\nabla_\mu D^\mu$, with $D^\mu$ being a local dilatation current. 
This signals a breaking of the global Weyl symmetry, 
meaning that the physical state of the theory will not be invariant under dilatations 
or special conformal transformations, but the local Weyl symmetry will in fact not be violated. 

Finally, one might worry that this unphysical dependence on $\mu$ changes the equations of motion for $\theta$. 
This is not the case however, since $\theta$ is a pure gauge field and, since the equations of motion are gauge 
independent,  once on-shell one can always go to the gauge $\theta = {\rm const}$. 
In this gauge, the Goldstone degree of freedom is incorporated in some 
other field~(which will typically be the metric tensor) and since the equations of motion are gauge independent the equation of motion for $\theta$ is 
just a constraint implied by the equations of motion of the field that has``eaten-up" the Golstone field.

\subsection{Higher order Ward identities}
\label{Higher order Ward identities}

To conclude, we shall list the higher order Ward identities that follow from invariance of the quantum theory under conformal
transformations. Some identities -- in particular the identities involving a three-point function --
can receive anomalous contributions from interacting
fermions. Although we do not analyse such contributions in detail here,
we are confident that the arguments provided in this paper apply also 
to conformal Ward identities for higher point functions.
Once again, this belief is motivated by the fact that quantisation on curved spacetimes
is Weyl invariant.

Consider the following $n$-point function of the theory,
\begin{equation}
\label{n-point.function} iG^{\{n\}}(x_1,\cdots,x_n) \equiv \langle T\left\{\phi(x_1)\cdots\phi(x_n)\right\}\rangle
=  \int\bar{\mathcal{D}}\phi  e^{iS[\phi]} \phi(x_1)\cdots\phi(x_n)\,,
\end{equation}
and performing a conformal transformation, $\phi\rightarrow (1+\gamma \omega)\phi$ we would find,
\begin{eqnarray}
\label{n-point.function.ward.identity}
\delta iG^{\{n\}}(x_1,\cdots,x_n) &=& \gamma \int_x\omega(x) \sum\limits_{i=1}^n \frac{\delta(x-x_i)}{\sqrt{-g(x_i)}}iG^{\{n\}}(x_1,\cdots,x_n) \\
&&\hskip-4cm =  i\int\bar{\mathcal{D}}\phi e^{iS[\phi]}
\int_x\bigg (-\frac{D-2}{2}\frac{1}{\sqrt{-g}}\frac{\delta S}{\delta \phi(x)} \omega(x) \phi(x)
-\frac{2}{\sqrt{-g}}\frac{\delta S}{\delta g^{\mu\nu}(x)} \omega(x) g^{\mu\nu}(x)
\label{n-point.function.ward.identity.2} \\
&+& \frac{1}{\sqrt{-g}}\frac{\delta S}{\delta \mathcal{T}_\mu(x)} \partial_\mu\omega(x)\bigg )\phi(x_1)\cdots\phi(x_n)\,,\nonumber
\end{eqnarray}
where $\int_x \equiv \int\text{d}^Dx\sqrt{-g}$, and $\gamma$ is the scaling dimension of the quantum field 
$\hat \phi$.
The first term in~(\ref{n-point.function.ward.identity.2}) vanishes due to the Ehrenfest theorem~(\ref{EoM.Time.Ordered.Product})
and hence -- up to a boundary term -- we are left with,
\begin{eqnarray}
\label{Ward.Identities.n.points.function}
 \Big\langle T\Big\{\big[\bar\nabla_\mu\Pi^\mu(x)+T_\mu^\mu(x)\big]\phi(x_1)\cdots\phi(x_n)\Big\}\Big\rangle
 = -\gamma\sum\limits_{i=1}^n \frac{\delta(x-x_i)}{\sqrt{-g(x_i)}}G^{\{n\}}(x_1,\cdots,x_n)
\,.
\end{eqnarray}
Upon expanding the time-ordering operator in~(\ref{Ward.Identities.n.points.function})
 and using the fact that $\delta(y^0-x^0)\left [ \Pi^0(y), \phi(x)\right ] =-[(D-2)/2] i \delta^D(x-y)$, 
%since the temporal component of the $\Pi^\mu$ current generates (canonical)~Weyl transformation, 
%can be used to show that 
it is easy to see that Eq.~(\ref{Ward.Identities.n.points.function}) can be rewritten as, 
\begin{eqnarray}
\label{Ward.Identities.n.points.function.anomalous.dimension}
&&\bar\nabla_\mu \Big\langle T\Big\{\Pi^\mu(x)\phi(x_1)\cdots\phi(x_n)\Big\}\Big\rangle + \Big\langle T\Big\{T_\mu^\mu(x)\phi(x_1)\cdots\phi(x_n)\Big\}\Big\rangle
 \\
 &&\hskip 3cm
= -\left(\gamma-\frac{D-2}{2}\right)\sum\limits_{i=1}^n \frac{\delta(x-x_i)}{\sqrt{-g(x_i)}}G^{\{n\}}(x_1,\cdots,x_n)\nonumber
\,.
\end{eqnarray}
The factor $\gamma-(D-2)/2$ on the right hand side of 
Eq.~(\ref{Ward.Identities.n.points.function.anomalous.dimension})
is just the anomalous dimension of $\hat \phi$.  

A proper understanding of the higher conformal Ward 
identities~(\ref{Ward.Identities.n.points.function.anomalous.dimension}) 
is important and we intend to consider their full ramifications 
for conformal anomalies in a separate publication.

%%%%%%%%%%%%%%%%%%%%%%%%%%%%%%%%%%%%%%%%%%%%%%%%%%%%%%%%%%%%%%%%%%%%%%%%
%%%%%%%%%%%%%%%%%    C O N C L U S I O N    %%% %%%%%%%%%%%%%%%%%%%%%
%%%%%%%%%%%%%%%%%%%%%%%%%%%%%%%%%%%%%%%%%%%%%%%%%%%%%%%%%%%%%%%%%%%%%%%%

\section{Conclusion}
\label{Conclusion}

A vast amount of literature agrees that conformal (Weyl) symmetry is a very special local symmetry in the sense that,
if it is symmetry of a classical theory, it generically gets broken by quantum effects.
It is interesting to note that all other local symmetries that are realised in Nature -- including gauge symmetries and diffeomorphisms --
are widely believed to be respected at the quantum level provided they are realised classically, 
with the notable exception of chiral symmetry.

In this work we argue that there is a way of preserving Weyl symmetry at the quantum level.
The trick is to add to the action a dilatation current (that is independent of the stress-energy tensor)
and a corresponding compensating Weyl field. In light of our former work~\cite{Lucat:2016eze}
(see section~\ref{Gravity with Torsion and its Symmetries} for a brief introduction into gravity theory 
with torsion)
the torsion trace appears as the natural candidate for the Weyl field since torsion makes gravity conformal
and in addition it is geometric and thus couples universally to all matter fields,
respecting thus the equivalence principle.

In fact the usual conformal anomalies found in literature can be 
reinterpreted as the anomalies associated with breaking of global scaling symmetry. Indeed, 
we argue in section~\ref{Ward identities for Weyl Symmetry} that 
a non-vanishing of the trace of the energy-momentum tensor,
$\langle T_\mu^\mu\rangle\neq 0$, is a telltale sign for a breakdown of global rescaling symmetry by quantum effects. 

On the other hand, the story of local Weyl symmetry is completely different. Indeed,  
in section~\ref{Ward identities for Weyl Symmetry} we show that local Weyl symmetry needs not get broken by quantum effects.
In particular there we show that, in presence of a compensating Weyl field $\mathcal{T}_\mu$, the conformal Ward identities get modified,
but remain satisfied.
For example, the fundamental conformal Ward identity gets modified to~(\ref{conformal.symmetry.Ward.identity}),
{\it i.e.} any anomalous contribution to the trace of the energy-momentum, $\langle T_\mu^\mu\rangle$,
is not interpreted as a violation of local Weyl symmetry, but instead it just sources divergence of 
the current $\Pi^\mu$, which acts as a source for the Weyl field and can be thus interpreted 
as the curved space generalisation of the dilatation current. Unlike in flat spacetimes however, where  
the dilatation current depends on the energy-momentum tensor, in curved spaces it generally does not because 
in curved spaces there exists an independent Weyl one-form which can be naturally identified 
with the torsion trace. 

We proceed and in~\ref{Conformal anomaly in an interacting scalar theory},
\ref{Conformal anomaly in Yukawa theory} 
and~\ref{Conformal anomaly in an interacting non-Abeliean gauge theory}
we consider several simple interacting theories of matter fields that are classically conformal 
in four spacetime dimensions
and show that they all satisfy the fundamental Ward identity~(\ref{conformal.symmetry.Ward.identity}).
This then lends strong support to the statement that, if the standard model is made classically conformal
(which can be done {\it e.g.} by replacing the Higgs mass term, $\mu^2 H^\dagger H$, by 
an interaction term with a scalar singlet $\phi$, $-\lambda_{\phi H}\phi^2H^\dagger H$), it will
remain conformal at the quantum level. The importance of this statement cannot be underestimated
for building conformal extensions of the standard model, in which the symmetry is respected 
in the ultraviolet, but `broken' by scalar condensate(s) mediated by quantum effects such as in 
the Coleman-Weinberg mechanism.  

Furthermore, we point out in~\ref{The role of scale mu}
that - even though quantisation procedures in general respect conformal symmetry
(as shown in~\ref{Weyl symmetry in the quantum theory})
 -- it is generically broken by all standard regularisation procedures. 
We first point out that,  in light of Weyl transformations,
scale dependence introduced by conventional regularisation schemes can be 
thought of as a constant scale in a particular conformal gauge 
(given by ${\cal T}_\mu^L\equiv\partial_\alpha \theta =0\implies \theta={\rm constant}$).
Next we show that, elevating the usual notion of spacetime constancy 
($\partial_\alpha\mu=0\implies \mu={\rm constant}$) to the suitable gauge indepedent notion
($\bar\nabla_\alpha\mu=0\implies {\cal T}_\alpha = -[2/(D\!-\!2)]\partial_\alpha\ln(\mu)$)
restores the conformal Ward identity~(\ref{conformal.symmetry.Ward.identity})
implied by any effective action. This is of course 
not a coincidence, as our procedure restores Weyl symmetry in regularisation procedures.
It is interesting to point out that one can pick the gauge $\mu ={\rm constant}$ at the level of equations
of motion (on-shell), such that regarding any physical quantity ${\cal O}_i[\mu]$ 
obtained by variation of an effective action,
there is no difference whatsoever between our procedure and conventional regularisation procedures. 
Indeed, since ${\cal O}_i[\mu]$  is an on-shell quantity, one can choose the scale $\mu$ in ${\cal O}_i[\mu]$ 
to be constant (which is what is usually done) 
and then set $\mu$ to a convenient physical quantity 
(such as an invariant energy ${\cal E}$ in scattering processes) to finally obtain
${\cal O}_i[{\cal E}]$. Such a procedure fully respects the conformal Ward 
identity~(\ref{conformal.symmetry.Ward.identity}).
It is worth pointing out that our proposal to modify standard regularisation schemes 
is supported by our consideration of a 
representative sample of interacting quantum field theories 
in sections~\ref{Conformal anomaly in an interacting scalar theory}--\ref{Conformal anomaly in an interacting non-Abeliean gauge theory} in the sense that
the operator methods and effective action approach yield equivalent conformal Ward identities.

The Gauss-Bonnet term ${\cal E}_4$~(\ref{Euler.density}) in section~\ref{Boundary (total derivative) terms}
is a geometric scalar that contributes to the conformal 
anomaly and deserves additional reflection. Namely, ${\cal E}_4$  
is topological in $D=4$, which means that it can be written as a derivative 
of a vector, ${\cal E}_4=\bar\nabla_\mu {\cal V}_4^\mu$. 
This then implies that the Gauss-Bonnet integral, 
which in Euclidean spaces gives the Euler characteristic of the manifold that can be represented 
as an alternating sum of the Betti numbers, is related to the topology of the spacetime manifold 
and therefore, ultimately, to the quantum state of the gravitational field. 
Then quantum states with different Betti numbers fall into distinct topological classes 
which might be of fundamental importance in 
specifying the vacuum state of the gravitational field and, {\it via} its connection to 
the dilatation current, it can lead to creation of particles.

In order to  get a better idea on what it all means, let us recall 
the well known the chiral anomaly in particle physics, which states that the chiral current
in the standard model is anomalous in the sense that its divergence is sourced by the Chern-Simons' density
(which is a pseudo-scalar proportional to the product of electric and magnetic fields). 
Then a change in the Chern-Simons' number (which is the spatial volume integral
of the Chern-Simons' density) signals creation of chiral fermions out of the Dirac sea. 
Analogously, a change in the Euler characteristic signifies creation of 
particles associated with the dilatation current $\Pi^\mu$, which 
are scalar particles. These particle deserve a name and can be called {\it conformalons}
or {\it weylons}.
 
While we show in section~\ref{Weyl symmetry in the quantum theory} 
that quantisation of simple constrained systems is conformal
-- see Appendix~A for a consideration of quantisation of an Abelian gauge theory -- we postpone the analysis of 
non-Abelian gauge theories and gravity to future work. 

Finally, for completeness in~\ref{Higher order Ward identities}
we show how to derive higher order conformal Ward identities, but 
leave to future work to rigorously prove that they are indeed satisfied.

\section*{Acknowledgements}

 This work supported in part by the D-ITP consortium, a program of
the Netherlands Organisation for Scientific Research (NWO) that is funded by the Dutch
Ministry of Education, Culture and Science (OCW). We acknowledge financial support from
an NWO-Graduate Program grant.

%%%%%%%%%%%%%%%%%%%%%%%%%%%%%%%%%%%%%%%%%%%%%%%%%%%%%%%%%%%%%%%%%%%%%%%%%%%%%%%%%%%%%%%%%%%%%%
%%%%%%%%%%%%%%  A P P E M D I X   A:  A B E L I A N   G A U G E   F I E L D S  %%%%%%%%%%%%%%%
%%%%%%%%%%%%%%%%%%%%%%%%%%%%%%%%%%%%%%%%%%%%%%%%%%%%%%%%%%%%%%%%%%%%%%%%%%%%%%%%%%%%%%%%%%%%%%

\section*{Appendix A: Abelian Gauge Theory}
\label{Abelian Gauge Theory}

Gauge theories and gravity constitute constrained systems and therefore their quantization requires a special attention.
For simplicity here we consider the symplest case: path integral quantization of an Abelian gauge theory, but we expect
analogous results to hold for non-Abelian gauge theories and gravity.

For a theory with constraints Dirac quantisation~\cite{Dirac:1964} is the method of choice.
In this procedure, instead of replacing Poisson brackets of quantities $A,B$ with commutators as it is
prescribed in canonical quantisation,
%
%\begin{equation}
$
\{A,B\}\rightarrow  [\hat A,\hat B]/(i\hbar)
%\frac{[\hat A,\hat B]}{i\hbar}
\, ,
$
%\label{canonical quantisation}
%\end{equation}
%
one constructs Dirac brackets $\{\cdot , \cdot\}_D$
and replaces them with commutators according to,
\begin{equation}
\{A,B,\}_D\rightarrow  \frac{[\hat A,\hat B]}{i\hbar}
\, .
\label{Dirac quantisation}
\end{equation}
Dirac brackets are constructed from an extended Hamiltonian, in which all (independent) constraints are
enforced by the introduction of Lagrange multipliers. The choice of Lagrange multipliers is in principle arbitrary, and the
freedom of their choice is known as gauge freedom. A particular choice of how Lagrange multipliers depend on fields and
their canonical momenta corresponds to a choice of gauge (this notion of gauge generalises the usual notion of gauge fixing
in gauge theories). One can then show that different gauge choices
yield identical answers for expectation (on-shell) values of Hermitean operators (physical observables).
This independence of gauge of on-shell quantities is then the precise sense in which gauge freedom exists.
For brevity in this Appendix we focus primarily on the path integral quantization and refer to
Ref.~\cite{Senjanovic:1976br}, in which it was proved that Dirac quantization and path integral quantization presented here
are equivalent in the sense that they yield identical in-out scattering amplitudes.

We begin our consideration by noting the classical action for an Abelian gauge field $A_\mu$,
\begin{eqnarray}
\label{Gauge.field.action}
S_{\rm EM} &=& \int \text{d}^Dx \sqrt{-g} \left [-\frac{1}{4}g^{\mu\lambda}g^{\nu\sigma} F_{\mu\nu} F_{\lambda\sigma}\right ] = \\
%\!&=&\! \int \text{d}^4x \sqrt{-g}-\frac{1}{4} F_{\mu\nu} F_{\lambda\sigma} \left [\left (g^{\mu\lambda} - \frac{n^\mu %n^\lambda}{\|n\|^2}\right ) +  \frac{n^\mu n^\lambda}{\|n\|^2}\right ]\left [\left (g^{\nu\sigma} - \frac{n^\nu n^\sigma}{\|n\|^2}\right %) +  \frac{n^\nu n^\sigma}{\|n\|^2}\right ] = \\
\!&=&\! \int \text{d}^4x \sqrt{-g}\left [-\frac{1}{4}
\left (g_\perp^{\mu\lambda} + g_\parallel^{\mu\lambda} \right )\left (g_\perp^{\nu\sigma} +  g_\parallel^{\nu\sigma}\right )
   F_{\mu\nu} F_{\lambda\sigma} \right ]
\,,
\end{eqnarray}
where $F_{\mu\nu}=\partial_\mu A_\nu-\partial_\nu A_\mu$ is the gauge field strength,
$g_\perp^{\mu\lambda} = g^{\mu\lambda}- \frac{n^\mu n^\lambda}{\|n\|^2}$ is the induced (inverse) metric on a space-like
hyper-surface $\Sigma$ ($n^\mu\perp\Sigma$) and $g_\parallel^{\mu\lambda}=\frac{n^\mu n^\lambda}{\|n\|^2}$.
The canonical momentum of $A_\mu$ is given by,
\begin{eqnarray}
\label{gauge.field.momentum}
\pi^\mu \!&=&\! \frac{\delta S}{\delta  (n^\nu\partial_\nu{A}_\mu(x))}
%  = - \sqrt{-g} \frac{n^\sigma}{\|n\|^2} \left (g^{\mu\lambda} - \frac{n^\mu n^\lambda}{\|n\|^2}\right ) F_{\sigma\lambda}=\\
 =- \sqrt{-g} \frac{n^\sigma}{\|n\|^2} g_\perp^{\mu\lambda}  F_{\sigma\lambda}
 \implies
    n^\mu F_{\mu\nu} =- \frac{\pi^\mu g_{\mu\nu}^\perp }{ \sqrt{-g} \|n\|^{-2}}
 \,\, \&\,\,
    n_\mu \pi^\mu = 0
 \,.\quad
 \label{time.momentum}
\end{eqnarray}
The last equation in Eq.~(\ref{time.momentum}) is a constraint.
When one chooses time such $\Sigma$ is a hypersurface of constant time, then  $n_\mu=\delta_\mu^0$
and the condition $n_\mu \pi^\mu = \pi^0 = 0$ sets the temporal momentum of the gauge field to zero.
Other (secondary) constraints are obtained by taking the commutator of~(\ref{time.momentum}) with the Hamiltonian,
which is obtained by taking a Legendre transform of the Lagrangian in~(\ref{Gauge.field.action}),
\begin{equation}
\begin{split}
\label{Hamiltonian.Gauge.Field}
H= \int \text{d}^{D-1}x \bigg ( -\frac{ g^\perp_{\mu\nu}\pi^\mu\pi^\nu}{2\sqrt{-g}\|n\|^{-2}}
 + \pi^\mu n^\nu\partial_\mu A_\nu
  + \frac{\sqrt{-g}}{4}g_\perp^{\alpha\gamma} g_\perp^{\beta\delta} F_{\alpha\beta} F_{\gamma\delta}
    +\lambda_0 \, n_\mu \pi^\mu\bigg )\,,
\end{split}
\end{equation}
where, in order to enforce the constraint~$n_\mu\pi^\mu=0$, we have introduced a Lagrange multiplier, $\lambda_0$.
Secondary constraints can be obtained by computing the Poisson bracket of the constraint $n_\mu\pi^\mu$ with the Hamiltonian.
Using $\{\pi^\mu(\vec{x}), A_\nu(\vec{y})\} =\delta^\mu_\nu \delta(\vec{x}-\vec{y})$, we find,
\begin{equation}
\label{Secondary.Constrain.Gauss.Law}\Phi= n_\sigma\left\{\pi^\sigma, H \right \} = - n_\sigma \partial_\mu\left (n^\sigma \pi^\mu\right )  -  \sqrt{-g}n_\beta \nabla_\alpha \left (g_\perp^{\alpha\gamma} g_\perp^{\beta\delta} F_{\gamma\delta}\right ) \,.
\end{equation}
One can then verify that no further independent constraints are generated by taking other Poisson brackets.
Note also that $\Phi$ is a covariant constraint: if the definition of the momentum $\pi^\mu$ is plugged in we get,
\begin{equation}
\begin{split}
\label{secondary.costrain.covariant.form}
\Phi =& - n_\rho \partial_\mu\left ( \sqrt{-g} \frac{n^\rho n^\sigma}{\|n\|^2} g_\perp^{\mu\lambda}  F_{\sigma\lambda}\right ) 
= - \sqrt{-g} \|n\| \frac{n_\rho}{ \|n\|}\bar{\nabla}{}_\mu\left ( F^{\mu\rho}\right ) 
\,.
\end{split}
\end{equation}
The physical meaning of $\Phi$ can be divulged/disclosed by making use of the Stokes' theorem.
Since the conformal dimension of the dual 2-form of the
gauge field strength, $\tilde{F}_{\alpha\beta} = \epsilon_{\alpha\beta}{}^{\mu\nu} F_{\mu\nu}$,
is $w = \frac{D-4}{2}$, we can
apply the conformal Stokes theorem~(\ref{stokes.theorem.2}) in $D=4$ to $\text{d}\tilde{F}$ on some subset of the spatial slice, $\mathcal{I}\subset\Sigma$,
\begin{equation}
\begin{split}
\label{stokes.theorem.application}  &\int_\mathcal{I} \bar{\nabla}_\lambda\tilde{F}_{\mu\nu} \text{d}x^\lambda\wedge\text{d}x^\mu\wedge\text{d}x^\nu=\int_\mathcal{I} \partial_\lambda\tilde{F}_{\mu\nu} \text{d}x^\lambda\wedge\text{d}x^\mu\wedge\text{d}x^\nu = \\
=&4\int_\mathcal{I}  \frac{n_\rho}{ \|n\|}\bar{\nabla}{}_\mu\left ( F^{\mu\rho}\right ) \sqrt{g^\perp}\text{d}^3x = 4\int_\mathcal{I} \Phi\,\text{d}^3x=\\
=&4\int_\mathcal{I}\left [\bar{\nabla}{}_\mu \left ( g^{\mu\nu}_\perp\frac{n^\alpha}{\|n\|} F_{\nu\alpha} \right )- F^{\lambda\rho} \bar{\nabla}_\lambda \left (\frac{n_\rho}{\|n\|}\right )\right ]\sqrt{g^\perp}\text{d}^3x \\
=& 4\int_\mathcal{I}\left [(g^\perp)^\mu_\lambda\bar{\nabla}{}_\mu \left ( g^{\lambda\nu}_\perp\frac{n^\alpha}{\|n\|} F_{\nu\alpha} \right )\right ]\sqrt{g^\perp}\text{d}^3x=\int_{\partial\mathcal{I}} {\tilde F}_{\mu\nu} \text{d}x^\mu\wedge\text{d}x^\nu
\,,
\end{split}
\end{equation}
where  we used the notation,
\begin{equation}
\frac{n_\rho}{\|n\|}\text{d}V=\frac{n_\rho}{\|n\|} \sqrt{-g^\perp} \text{d}x\text{d}y\text{d}z = \frac{1}{3!} \epsilon_{\rho\alpha\beta\gamma}\text{d}x^\alpha\wedge\text{d}x^\beta\wedge\text{d}x^\delta
\end{equation}
to denote the induced volume form on the surface $\mathcal{I} \subset\Sigma$, which transforms as a vector of scaling dimension $4$ under Weyl transformations. Furthermore, $\sqrt{g^\perp}\text{d}^3x = \sqrt{-g} \|n\|^{-1}\text{d}^3x$ is the induced volume form on $\Sigma$.
From Eq.~(\ref{stokes.theorem.application}) we see that the last equality represents
the electric flux associated with the field strength $F_{\mu\nu}$ through the two-dimensional surface $\partial\mathcal{I}$.

In presence of a source the constraint~(\ref{Secondary.Constrain.Gauss.Law}) gets modified from $\Phi=0$  to $\Phi= n_\mu J_\gamma^\mu$, where $J_\gamma^\mu$ is the electromagnetic current. This then modifies~(\ref{stokes.theorem.application}) to,
\begin{equation}
\label{Physical.meaning.constraint}
\int_{\partial\mathcal{I}} {\tilde F}_{\mu\nu} \text{d}x^\mu\wedge\text{d}x^\nu = \int_\mathcal{I} \frac{n_\mu}{\|n\|} J^\mu_\gamma \sqrt{g^\perp}\text{d}^3x\,.
\end{equation}
Hence Eq.~(\ref{Physical.meaning.constraint}) is the curved space
generalization of the Gauss' law, $ \int_{\partial\mathcal{I}} \vec{E}\cdot \text{d}\vec{S} = \int_{\mathcal{I}} \rho \text{d}V$.
In Eq.~(\ref{Physical.meaning.constraint}) the role of the electric field is played by the spatial vector, $g^{\mu\nu}_\perp F_{\nu\alpha}\frac{n^\alpha}{\|n\|}$, which on flat spaces with flat foliation results in $F^{i0}$.

The gauge condition that we can associate with the constraint $n_\mu \pi^\mu=0$ is naturally $n^\mu A_\mu$, while the one to associate with $n_\sigma\partial_\mu \left (n^\sigma\pi^\mu\right) = \sqrt{g^\perp}(g^\perp)^\mu_\nu \bar\nabla_\mu \frac{\pi^\nu}{\sqrt{g^\perp}}$ should be, in light of the previous observations,
$(g^\perp)_\alpha^\beta\bar{\nabla}_\beta \left (A_\nu g^{\alpha\nu}_\perp\right )$.
Thus we can construct the path integral representation of the scattering amplitude
 in the gauge, $n^\mu A_\mu=0$, $\bar{\nabla}^\perp_\mu A^\mu_\perp = 0$. The Poisson brackets of
 the constraints with their associated canonical variables are,
\begin{equation}
\begin{pmatrix}
\{n_\mu\pi^\mu, n^\nu A_\nu\} &0\\
0& \{ \sqrt{g^\perp}(g^\perp)^\mu_\nu \bar\nabla_\mu \frac{\pi^\nu}{\sqrt{g^\perp}},\bar{\nabla}{}^\perp_\nu A^\nu_\perp\}
\end{pmatrix} = \begin{pmatrix}
\|n\|^2 \delta^{D-1}(\vec{x}-\vec{y})&0\\
0&\bar{\Box}{}^\perp\delta^{D-1}(\vec{x}-\vec{y})
\end{pmatrix}\,,
\end{equation}
where 
\begin{equation}
\bar{\Box}^\perp \delta^{D-1}(\vec{x}-\vec{y})= \frac{1}{\sqrt{g^\perp}}\left (\partial_\mu - \left (\frac{D-2}{2}\right )T_\mu\right ) \sqrt{g^\perp} g^{\mu\nu}_\perp\partial_\nu \delta^{D-1}(\vec{x}-\vec{y})\,,
\end{equation}
transforms covariantly under Weyl rescalings in $D=4$, acting on a function with scaling dimension $0$. 

Following~\cite{Senjanovic:1976br} we can now write the path integral representation of the in-out scattering matrix
as,~\footnote{Even though the path integral~(\ref{Photon.Field.Path.integral} )
 is derived for the free theory, its expression is valid also in interacting theories.
  The constraints in interacting theories get modified in the sense that they involve sources. This modification
   does not change the invariance of the path integral measure
  under gauge or conformal transformations, but might introduce non-trivial field dependence in the determinant factor,~$\det\left( \bar{\nabla}{}_i\partial^i \delta(\vec{x}-\vec{y})\right)$, since some constraints and gauge conditions may not commute with each other.}
\begin{equation}
\begin{split}
\label{Photon.Field.Path.integral}
\langle {\it in}|{\it out}\rangle = &\int \prod\limits_\sigma \mathcal{D}\pi^\sigma\mathcal{D}A_\sigma  \,\delta\left( n_\mu\pi^\mu \right ) \delta\left(n^\nu A_\nu \right ) \delta\left(  \partial_\alpha\pi^\alpha_\perp\right ) \delta\left(\bar{\nabla}{}^\perp_\beta A^\beta_\perp \right )\left|\det\left(\|n\|^2\delta^{D-1}(\vec{x}-\vec{y})\right)\right |\\
 &\times\left|\det\left(\bar{\Box}{}^\perp \delta^{D-1}(\vec{x}-\vec{y})\right)\right | \exp\left(i\int\text{d}^Dx \big[\pi^\mu n^\nu\partial_\nu A_\mu - H\big]\right ) \,.
 \end{split}
\end{equation}
In Ref.~\cite{Senjanovic:1976br} it was shown that the scattering amplitude~(\ref{Photon.Field.Path.integral}) is independent
on the gauge condition chosen. In addition Eq.~(\ref{Photon.Field.Path.integral})
is manifestly covariant (diffeomorphism invariant) and conformal (Weyl invariant), as can be easily checked.

%%%%%%%%%%%%%%%%%%%%%%%%%%%%%%%%%%%%%%%%%%%%%%%%%%%%%%%%%%%%%%%%%%%%%%%%%%%%%%%%%%%%%%%%
%%%%%%%%%%%  A P P E N D I X   B:  E H R E N F E S T   T H E O R E M    %%%%%%%%%%%%%%%%
%%%%%%%%%%%%%%%%%%%%%%%%%%%%%%%%%%%%%%%%%%%%%%%%%%%%%%%%%%%%%%%%%%%%%%%%%%%%%%%%%%%%%%%%

\section*{Appendix B: Ehrenfest Theorem}
\label{Appendix B: Ehrenfest Theorem}

In this appendix we derive the Ehrenfest theorem, which can be reinterpreted as
the Ward identity for infinitesimal fields translations. This theorem is useful for proving
the fundamental conformal Ward identity in section~\ref{Fundamental Ward identity}.
Consider now the time ordered product of $n$-fields, represented by the path integral,
\begin{equation}
\label{Time.Ordered.Product.1}
\langle {\it in}|T\{\phi(x_1)\cdots\phi(x_n)\}|{\it out}\rangle
= \frac{1}{\langle {\it in}|{\it out}\rangle} \int  \bar{\mathcal{D}}\phi e^{i\int \text{d}^Dx \mathcal{L}_\phi} \phi(x_1)\cdots\phi(x_n)\,.
\end{equation}
The path integral~(\ref{Time.Ordered.Product.1}) is invariant under the field independent local shifts,
$\phi(x) \rightarrow \phi'(x) = \phi(x) + \xi(x)$, for which one can then write,
$ \langle {\it in}|T\{\phi(x_1)\cdots\phi(x_n)\}|{\it out}\rangle =  \langle {\it in}|T\{\phi'(x_1)\cdots\phi'(x_n)\}|{\it out}\rangle$, from which it follows that~\cite{Peskin:1995ev},
\begin{equation}
\begin{split}
\label{EoM.Time.Ordered.Product}
& \left\langle {\it in}\bigg|T\left\{\left (\frac{\delta}{\delta \phi(x)}\int\text{d}^Dx \mathcal{L}_\phi\right )\phi(x_1)\cdots\phi(x_n)\right\}\bigg|{\it out}\right \rangle = 0
\,,
%=&i \sum\limits_{i=1}^n\left\langle T\left\{\phi(x_1) \cdots \delta^D(x-x_i)\cdots\phi(x_n)\right\}\right\rangle \,.
\end{split}
\end{equation}
where any derivative operator acts inside the time-ordered product~\cite{Peskin:1995ev}.
When, on the other hand, the derivatives are pulled outside of the time-ordered product, one gets,
%In Eq.~(\ref{EoM.Time.Ordered.Product}) any derivative operator ought to be placed outside the time ordered %product~\cite{Peskin:1995ev},
%
\begin{equation}
\begin{split}
&\Box_x \left\langle {\it in}\bigg|T\left\{\phi(x)\phi(x_1)\cdots\phi(x_n)\right\}\bigg|{\it out}\right \rangle +   \left\langle {\it in}\bigg|T\left\{\left (\frac{\partial \mathcal{L}_{\it int}}{\partial\phi(x)}\right )\phi(x_1)\cdots\phi(x_n)\right\}\bigg|{\it out}\right \rangle= \\
&\hskip 6.7cm
 = i \sum\limits_{i=1}^n\left\langle T\left\{\phi(x_1) \cdots \delta^D(x-x_i)\cdots\phi(x_n)\right\}\right\rangle \,,
\end{split}
\label{Time.Ordered.Product.2}
\end{equation}
where we have assumed that the Lagrangian density can be split into the free part (${\cal L}_0$) and the part that contains
(polynomial) interactions (${\cal L}_{\rm int}$) as,
\begin{equation}
   {\cal L}_\phi = {\cal L}_0 + {\cal L}_{\rm int}
\,,\qquad
{\cal L}_0 = \frac12 g^{\mu\nu}\big(\nabla_\mu\phi\big)\big(\nabla_\nu\phi\big)
\,.
\label{lagrangian: split}
\end{equation}
The non-vanishing right hand side in~(\ref{Time.Ordered.Product.2}) is due to the non-commuting of
the time ordering operation and the derivative operators appearing
in the Lagrangian~(\ref{lagrangian: split}).~\footnote{Equation~(\ref{Time.Ordered.Product.2}) can be
proven by noting that one can rewrite time ordering operation as,
\[\begin{split}\left\langle {\it in}\bigg|T\left\{\phi(x_1)\cdots\phi(x_n)\right\}\bigg|{\it out}\right \rangle  = \sum\limits_{\sigma \in S_n} &\theta\left(x^0_{\sigma(1)}-x^0_{\sigma(2)}\right)\theta\left(x^0_{\sigma(2)}-x^0_{\sigma(3)}\right)\cdots \theta\left(x^0_{\sigma(n-1)}-x^0_{\sigma(n)}\right)\times \\
&\times\left\langle {\it in}\bigg|\phi\left(x_{\sigma(1)}\right)\cdots\phi\left(x_{\sigma(n)}\right)\bigg|{\it out}\right \rangle\,,\end{split}\]
where $S_n$ is the permutation group. Then, Eq.~(\ref{EoM.Time.Ordered.Product}) results in a series of equal time commutators, which reproduce the Dirac delta functions on the right hand side of~(\ref{Time.Ordered.Product.2}).}

In particular, Eq.~(\ref{Time.Ordered.Product.1}) implies the identity,
\begin{equation}
\label{Lemma.Two.Points.Functions} \left\langle {\it in}\bigg|T\left\{\left (\frac{\delta S}{\delta \phi(x)}\right )\phi(y)\right\}\bigg|{\it out}\right \rangle = 0\,,
\end{equation}
where all the derivatives act inside of the time-ordered product.
This equation -- also known as the Ehrenfest theorem -- is just the statement that the field operator satisfies
its equations of motion multiplied by one (or more) field(s),
upon time ordering and expectation values are exacted.

%%%%%%%%%%%% B I B L I O G R A P H Y %%%%%%%%%%%%%%%%%%%%%

\end{document}